\documentclass[aps,prc,twocolumn,floatfix,superscriptaddress,nofootinbib,preprintnumbers,longbibliography,amsmath,amsthm,amssymb]{revtex4-2}
\usepackage{graphicx}
\usepackage{amsfonts}
\usepackage{color}
\usepackage[colorlinks=true,linkcolor=blue,citecolor=blue]{hyperref}
\usepackage{dcolumn}% Align table columns on decimal point
\usepackage{bm}% bold math
\usepackage{braket}
\usepackage[normalem]{ulem}  % \sout{old text} for strikeout

\begin{document}
\allowdisplaybreaks[1]
\title{Fission Fragment Yields Of $^{235}$U$(n_{th},f)$ Evaluated By The CCONE Code System}

\author{Futoshi Minato}
\email[E-mail: ]{minato.futoshi.009@m.kyushu-u.ac.jp}
\affiliation{Department of Physics, Kyushu University, Fukuoka 819-0395, Japan}
\affiliation{RIKEN Nishina Center for Accelerator-Based Science, Wako, Saitama 351-0198, Japan}
\affiliation{Nuclear Data Center, Japan Atomic Energy Agency, Tokai, Ibaraki 319-1195, Japan}

\author{Osamu Iwamoto} 
%\email[E-mail: ]{}
\affiliation{Nuclear Data Center, Japan Atomic Energy Agency, Tokai, Ibaraki 319-1195, Japan}

%
%\preprint{KUNS-****}
\date{\today}% It is always \today, today,
\begin{abstract}
Fission fragment yield evaluations are one of the important nuclear data studies.
Fission accompanies various physical observables such as prompt fission neutron, prompt fission gamma, and delayed-neutrons.
When evaluating fission fragment yields, a study including correlations among those observables is essentially required.
However, fission fragment yield data in the past JENDL libraries have been made by focusing only on experimental fragment yields, decay heats, and delayed neutron yields, and they have not been expanded into a wider range of fission observables.
This is because the evaluation method adopted in the JENDL libraries could not study fission fragment yields and particle emissions from fragments simultaneously.
To solve this problem, a calculation system with CCONE code is newly developed to estimate not only independent and cumulative fission fragment yields but also prompt fission neutron, prompt fission gamma, decay heats, and delayed-neutrons simultaneously.
This system enables us to study a correlation between various fission observables.
To determine lots of parameters in this system efficiently, a Gaussian process and a least square fitting are adopted.
We tested the calculation system through a thermal neutron-induced fission on $^{235}$U.
In this paper, we demonstrate the performance of the parameter search method and show that experimental fission fragment yield data and other observables resulting from fission are reproduced well by the new calculation system.
\end{abstract}
\maketitle
\section{Introduction}
\label{sect:Intro}
Distributions of fission fragments resulted from neutron-induced fission are one of the key data for developments of next-generation nuclear reactors and disposals of spent nuclear fuels.
Accurate data of fission fragment yields are also required to monitor active reactors with reactor neutrinos~\cite{RevModPhys.92.011003}.
Evaluated data of fission fragment yields are currently available in several nuclear data library such as Japanese Evaluated Nuclear Data Library (JENDL), Evaluated Nuclear Data File (ENDF), and Joint Evaluated Fission and Fusion (JEFF).
\par
Recently, world-wide activities that evaluate new fission fragment yields based on recent knowledge on nuclear physics gains the momentum~\cite{osti_1807955}.
The movements are attributed from the background that discrepancies from experimental data are obtained for fission observables such as independent fission fragment yields (IFY), and spectra and multiplicities of prompt fission neutron (PFN) and prompt fission $\gamma$-rays (PFG) when calculating them with evaluated data of fission fragment yields.
The origin of this problems is simply due to that conventional approaches of fission fragment yield evaluations adopted in JENDL, ENDF, and JEFF have not respected all observables resulted from fission in the evaluations.
One of the typical examples is an overestimation of delayed-neutron yields in JENDL-4.0 calculated with a summation method of the fission product yield and decay data~\cite{Shibata2012}.
Although this problem was drastically mitigated in the latest version of JENDL, JENDL-5~\cite{JENDL-5}, the fission fragment yield data are still evaluated independently from PFN, PFG, and so on.
To provide versatile fission fragment yield data, it is required to comprehensively consider various observables resulted from fission besides its fragment yields.
\par
Neutron-induced fission is a complicated dynamics that involves energy dissipation, many deformation degrees of freedom, fluctuations along fission path, and so on.
To fully understand them, microscopic theoretical models that consider nuclear correlations among them are becoming a powerful tool.
However, those approaches still do not reach a satisfactory level for practical applications for nuclear data evaluations.
Recent macroscopic approaches for fission dynamics based on a Langevin models show a good agreement with experimental data of pre-neutron fission fragment yields rather than microscopic models~\cite{Huang2022, Tanaka2023, Fujio2023}.
However, those approaches also fail to reproduce pre-neutron fission fragment yields and total kinetic energy (TKE) distributions simultaneously, which become crucial to calculate particle evaporation from fragments, at a low incident neutron energy.
For these reasons, at least for pre-neutron fission fragment yields and TKE data, a phenomenological approach is still the most reliable if experimental data enough to adjust the parameters are available.
Once we obtain pre-neutron fission fragment yields and TKE data, we can calculate various fission observables with a particle evaporation model.
\par
We recently developed a new system of fission fragment yield evaluation using the Comprehensive COde for Nuclear data Evaluation~(CCONE)~\cite{Iwamoto2016}.
By using a phenomenologically determined pre-neutron fission fragment and TKE data, it calculates not only independent and cumulative fission yields, but also various observables resulting from fission using with decay data and the Hauser-Feshbach statistical model that describes particle evaporation from compound states.
The basic idea has already been applied in some studies using with BeoH code~\cite{Okumura2018, Lovell2021, Okumura2021}, FIFRELIN code~\cite{Litaize2015, Piau2023}, and FREYA~\cite{Verbeke2015, Verbeke2018}.
The characteristic point of our system is a new feature in the parameter search.
Since fission calculations involves a lot of parameters, it is hard to determine them without much time cost.
To overcome this issue, we used an approach combining a Bayesian optimization with Gaussian process (hereafter, we simply call "GP") and a generalized least squares (GLS) methods.
Multiplying the advantages of GP and GLS, an appropriate parameter set can be sought rather efficiently.
We tested our system by a thermal neutron-induced fission on $^{235}$U.
Sect.~\ref{sect:II} describes the framework to evaluate the fission fragment yields with the newly developed CCONE system.
Sect.~\ref{sect:3} demonstrates the result of calculations and discuss the difference from JENDL-5.
Sect.~\ref{sect:summary} summarize this work and give our perspectives.
\section{Methodology}
\label{sect:II}
\begin{figure*}[tb]
\centering
\includegraphics[width=0.45\linewidth, bb=280 170 700 515]{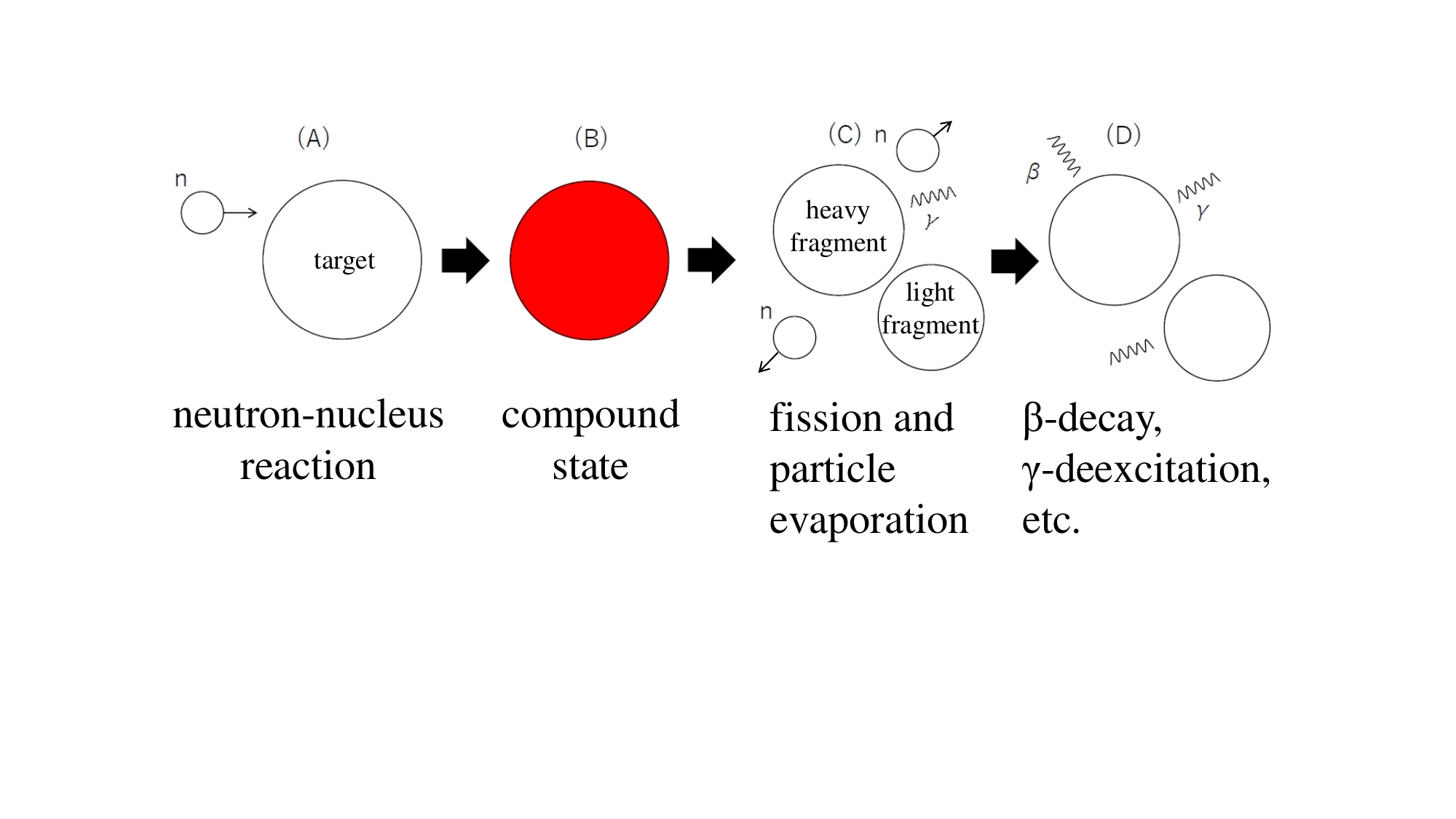}
\caption{Schematic picture of CCONE system of the fission fragment yield calculation.}
\label{fig:schematic}
\end{figure*}

The calculation of fission fragment yield of our system with CCONE is separated to four parts.
Figure~\ref{fig:schematic} illustrates each part schematically.
Part (A) and (B) in Fig.~\ref{fig:schematic} are the neutron-nucleus reaction and the formation of compound nucleus, respectively.
The calculation of these parts is carried out with a coupled-channel optical model, distorted-wave born approximation, and two-component exciton model that are already included in the former CCONE~\cite{Iwamoto2016}.
The detail of the calculation is described in Refs.~\cite{Iwamoto2007, Iwamoto2016}, and we do not go into the details in this paper.
The newly implemented modules in the present CCONE are part (C) and (D) that are particle evaporations from pre-neutron fission fragments and $\beta$ decays of post-neutron fission fragments toward stable nuclei, respectively.
The particle evaporation process in part (C) is studied within a Hauser-Feshbach statistical model implemented in CCONE, while the $\beta$-decay process in part (D) is calculated with a Batemann approach code~\cite{OYAK} and the evaluated decay data of JENDL-5~\cite{JENDL-5}.
Pre-neutron mass yields are approximated by a sum of five Gaussians and charge distributions are estimated by the $Z_{p}$ model~\cite{Wahl2002} with parameter modifications.
Excitation energies of fission fragments are calculated by an assumption of anisothermal model~\cite{Kawano2013}.
In the following sections, we demonstrate the present approaches for part (C) and (D) in detail.
\subsection{Pre-neutron mass yields and TKE distributions}
When a fission channel is energetically open, CCONE reads data of pre-neutron fission mass yields and TKE defined as $Y(A)$ and $\mathrm{TKE}(A)$, respectively.
The same systematics as Refs.~\cite{Okumura2018, Lovell2021, Okumura2021} are adopted for $Y(A)$ and $\mathrm{TKE}(A)$ in this work, in which $Y(A)$ is calculated with a sum of five Gaussians 
\begin{equation}
    Y(A)=G_{0}(A,E_{in})+G_{1}(A,E_{in})+G_{2}(A,E_{in}),
\end{equation}
where
\begin{equation}
    G_{0}(A,E_{in})=\frac{W_{0}(E_{in})}{\sqrt{2\pi}\sigma(E_{in})}\exp\left(-\frac{(A-\mu_{0}(E_{in}))^{2}}{2\sigma_{0}^{2}(E_{in})}\right),
\end{equation}
\begin{equation}
\begin{split}
    G_{i}(A,E_{in})
    &=\frac{W_{i}(E_{in})}{\sqrt{2\pi}\sigma(E_{in})}
    \left\{
    \exp\left(-\frac{(A-\mu_{i}(E_{in}))^{2}}{2\sigma_{i}^{2}(E_{in})}\right)\right.\\
    &\left.+\exp\left(-\frac{(A-(A_{c}-\mu_{i}(E_{in})))^{2}}{2\sigma_{i}^{2}(E_{in})}\right)
    \right\},
\end{split}
\end{equation}
for $i=1,2$ and $\mathrm{TKE}$ is estimated from
\begin{equation}
\begin{split}
{\rm TKE}(A_{h})&=\left(p_{0}-p_{1}A_{h}\right)\left(1-p_{2}\exp\left(-\frac{\left(A_{h}-\frac{A_{c}}{2}\right)^{2}}{p_{3}}\right)\right)\\
&+\epsilon_{{\rm TKE}}.
\end{split}
\label{eq:tke}
\end{equation}
Here, $W_{i}, \sigma_{i}$ ($i=0, 1, 2)$ and $p_{i}$ $(i=0,1,2,3)$ are the parameters determined so as to reproduce experimental data, and $A_{h}$ and $A_{c}$ are the mass of a heavy fragment and a compound nucleus, respectively.
Incident neutron energy is set to be thermal neutron as $E_{in}=E_{\mathrm{thermal}}=0.0253$ eV. 
The last term $\epsilon_{\mathrm{TKE}}$ in Eq.~\eqref{eq:tke} is introduced so that the average value with fragment yields becomes equal to $\overline{\mathrm{TKE}}=171.1$ MeV, which is obtained from experimental data~\cite{Meadows1982, Madland2006, Duke2014}.
TKE of a fragment with mass $A$ and atomic number $Z$ is estimated by
\begin{equation}
{\rm TKE}(A_{l},Z_{l})={\rm TKE}(A_{h},Z_{h})={\rm TKE}(A_{h}) \frac{ Z_{l}Z_{h}}{N_{ZZ}(A_{h})},
\end{equation}
where the normalization coefficient denoted as $N_{ZZ}$ is defined as
\begin{equation}
N_{ZZ}(A_{h})=\frac{\displaystyle \sum_{l,h} Z_{l}Z_{h}C(A_{h},Z_{h})}{\displaystyle\sum_{(l),h}C(A_{h},Z_{h})}.
\end{equation}
Here, $l, h$ take all possible fragment pairs with mass $A_{l}$ and $A_{h}$ and 
$C(A,Z)$ is the charge distributions described in Sect.~\ref{sect:charge}
As introduced in Ref.~\cite{Okumura2018}, we also consider a fluctuation of TKE from Eq.~\eqref{eq:tke} by considering a width parameter
\begin{equation}
\sigma_{\mathrm{TKE}}(A_{l})=\sigma_{\mathrm{TKE}}(A_{h})=s_{0}-s_{1}\exp\left[-s_{2}\left(A_{h}-\frac{A_{c}}{2
}\right)^{2}\right],
\end{equation}
where $s_{i}$ $(i=0,1,2)$ are the parameters. 
This work adopted the same parameters as Ref.~\cite{Lovell2021} for those of the five Gaussians, $p_{i}$ and $s_{i}$.
%In actual calculation, TKE distributions of fragments with $\sigma_{\mathrm{TKE}}$ are approximated by
%
%\begin{equation}
%    P_{\mathrm{TKE}}(\mathrm{TKE},A,Z)=\frac{1}{\sqrt{2\pi\sigma_{\mathrm{TKE}}^{2}}}
%    \exp\left(-\frac{(\mathrm{TKE}-\mathrm{TKE}(A,Z))^{2}}{2\sigma_{\mathrm{TKE}}^{2}(A)}\right).
%\end{equation}
%
%and is used later to calculate excitation energies of fission fragments in sec.~\ref{sec:excitation}.
%\begin{equation}
%P_{\mathrm{TKE}}(\mathrm{TKE}, A, Z)=\frac{1}{\sqrt{2\pi}\sigma_{\mathrm{TKE}}}\exp\left(-\frac{TKE-\mathrm{TKE}(A,Z)}{2\sigma_{\mathrm{TKE}}^{2}(A)}\right)    
%\end{equation}
%
%
%
\subsection{Charge distributions}
\label{sect:charge}
For charge distribution, we used $Z_{p}$ model proposed by Wahl~\cite{Wahl2002}.
The pre-neutron fission fragment yields are represented by
\begin{equation}
    Y(A,Z)=Y(A) \times C(A,Z),
\end{equation}
where $C(A,Z)$ is given by
\begin{equation}
    C(A,Z)=\int_{Z-0.5}^{Z+0.5}\frac{1}{\sqrt{2\pi}\sigma_{p}(A)}\exp\left(-\frac{(Z'-Z_{p}(A))^2}{2\sigma_{p}(A)^{2}}\right) dZ'
    \label{eq:charge}
\end{equation}
\label{sect:II.b}
\begin{figure}
    \centering
    \includegraphics[width=0.98\linewidth, bb=0 20 620 490]{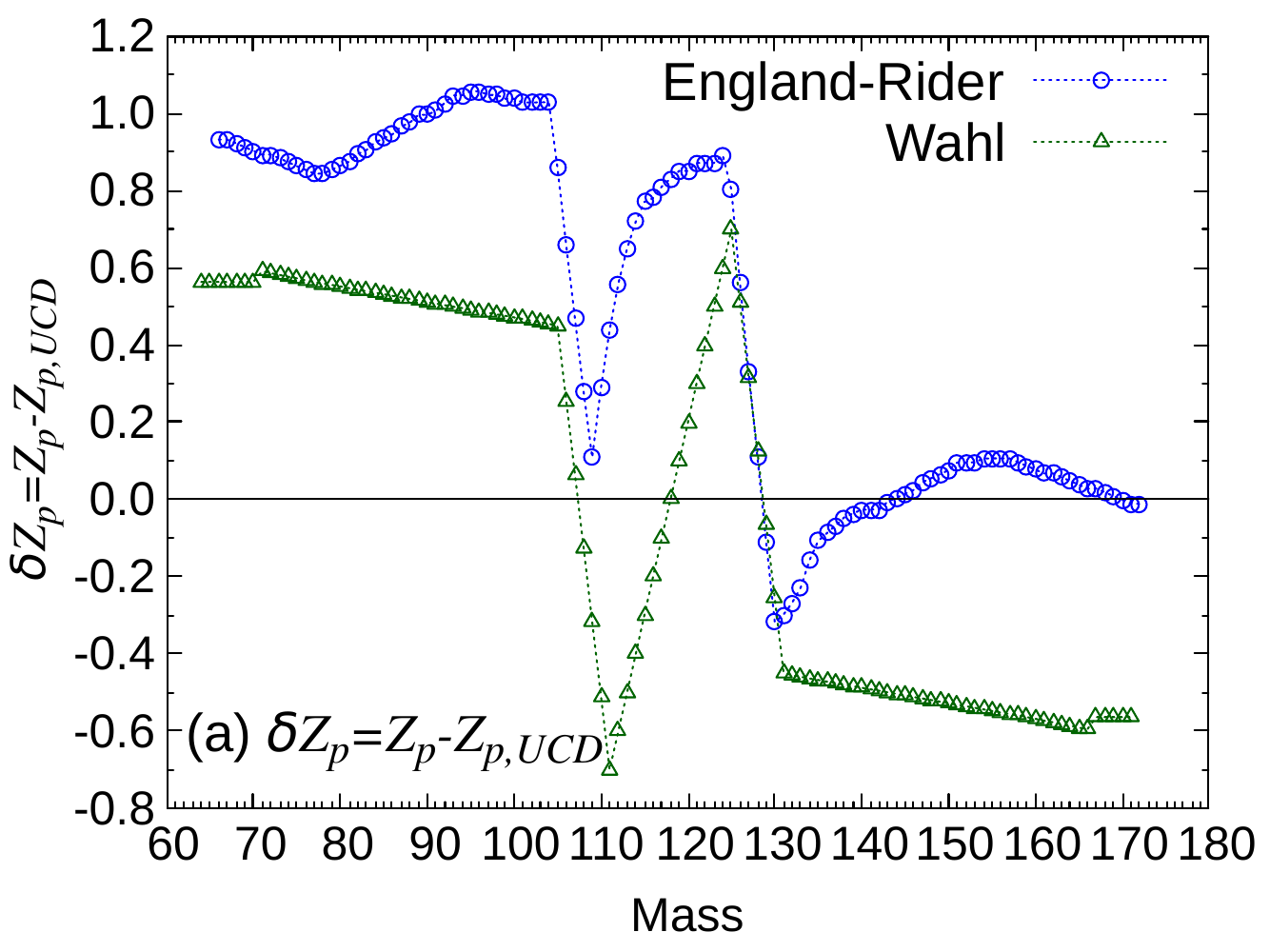}
    \includegraphics[width=0.98\linewidth, bb=0 20 620 490]{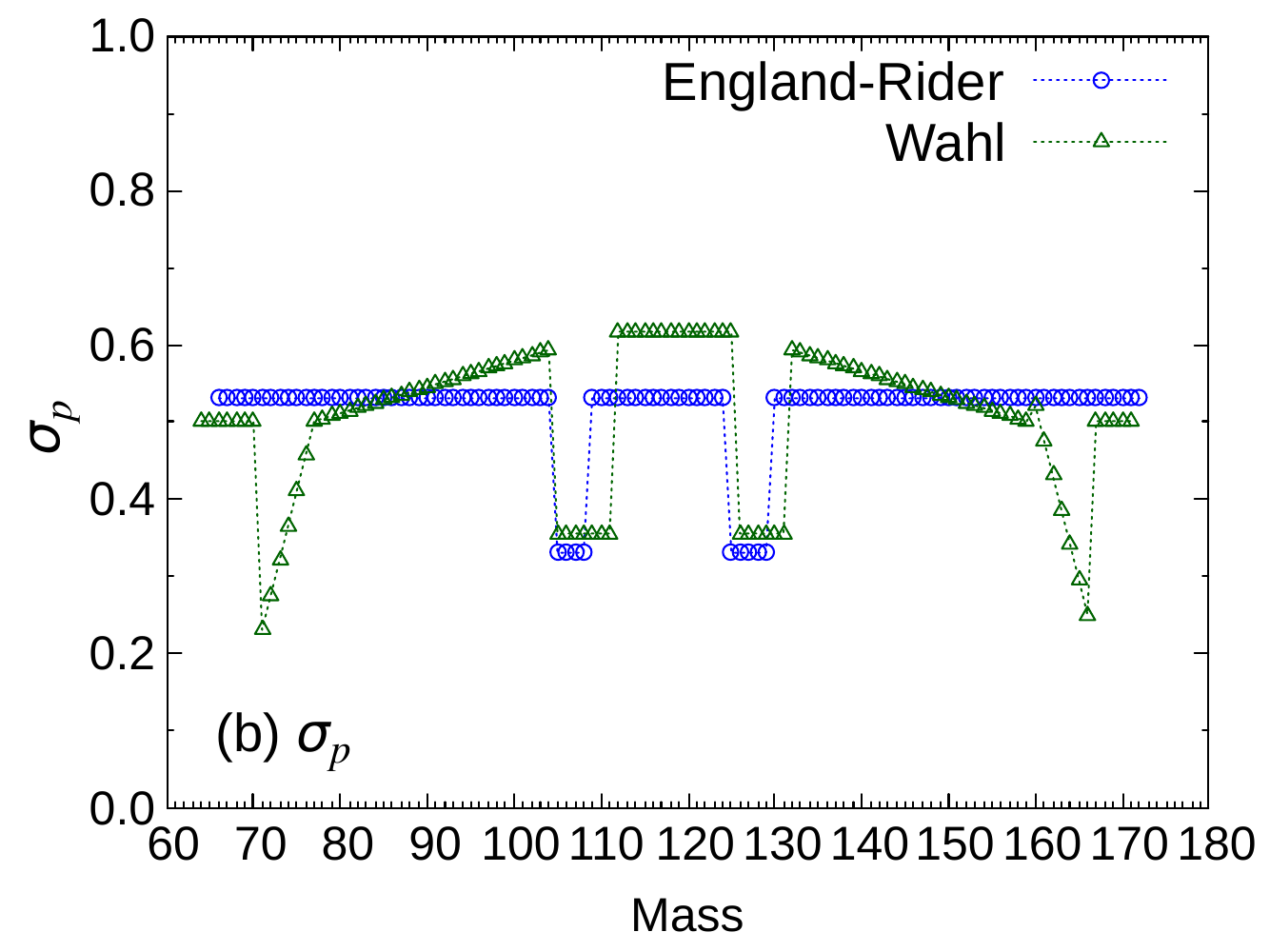}
    \caption{(a) Differences between unchanged charge distribution and evaluated value of the most probable proton number $Z_{p}$ of Wahl~\cite{Wahl2002} without the odd-even effect. For comparison, those evaluated by England and Rider~\cite{ER1993} in case of independent fission yields are shown. (b) Width parameter $\sigma_{p}$ of charge distributions.}
    \label{fig:Zp}
\end{figure}
In this work, $Z_{p}(A)$ and $\sigma_{p}(A)$ are approximated by
\begin{equation}
Z_{p}(A)=Z_{p, \mathrm{UCD}}(A)+\delta Z_{p}(A),
\end{equation}
and
\begin{equation}
\sigma_{p}(A_{l})=\sigma_{p}(236-A_{h})=0.50+\delta \sigma_{p}(A_{l}),
\end{equation}
respectively, where $\displaystyle Z_{p,\mathrm{UCD}}(A)\equiv\frac{Z_{f}}{A_{f}}A$ is the unchanged charge distribution, and $\delta Z_{p}(A)$ and $\delta \sigma_{p}(A)$ are free parameters to be determined by a parameter search method described later.
Here, $Z_{f}=92$ and $A_{f}=236$ are the atomic and mass numbers of fissioning compound nucleus, that is $^{236}$U.
We consider $Z_{p}(A)$ in the range of $64\le A \le 172$ as Wahl's evaluation~\cite{Wahl2002}, assuming $\delta Z_{p}(A_{h})=-\delta Z_{p}(A_{l})$ and $\delta \sigma_{p}(A_{h})=\delta \sigma_{p}(A_{l})$.
We then have $55$ free parameters for $\delta Z_{p}(A)$ and $\delta \sigma_{p}(A)$ from $A=64$ to $118$, respectively.
However, the symmetric fission is rare event for the thermal neutron-induced fission on $^{235}$U and $\delta Z_{p}(A=118)$ is not uniquely determined within the least square fitting, so that we set $\delta Z_{p}(A=118)=\delta \sigma_{p}(A=118)=0$ and excluded from the objects of parameter search.
\par
Figure~\ref{fig:Zp} shows $\delta Z_{p}(A)=Z_{p}-Z_{p,\mathrm{UCD}}$ and $\sigma_{p}(A)$ of the evaluated data of England and Rider~\cite{ER1993} and Wahl~\cite{Wahl2002}. 
Note that the evaluated data of England and Rider are for post-neutron fragment yields, while those of Wahl are for pre-neutron fragment yields.
Therefore, the evaluated data of England Rider are systematically larger than those of Wahl and rotationally asymmetric around the point of $(A=118,Z_{p}-Z_{p,\mathrm{UCD}}=0)$.
In Wahl's evaluation, $Z_{p}(A)$ are determined by separating them into four parts, which are far-wing $(A_{l} \le 70)$, wing ($71 \le A_{l} \le 77$), peak ($78 \le A_{l} \le 105$), and near-symmetry regions ($106 \le A_{l}\le 118$).
We can also observe in England and Rider evaluation that the $Z_{p}-Z_{p,\mathrm{UCD}}$ have different behavior at $A_{l}<78$, $78\le A_{l} < 96$, $96\le A_{l} < 104$, $104\le A_{l} <108$, and $108\le A_{l} < 118$.
In the following section, we demonstrate a performance of parameter search method by dividing $\delta Z_{p}(A)$ and $\delta \sigma_{p}(A)$ into $n=5$ parts, and then discuss the result of full parameters.
\subsection{Excitation energy distributions}
\label{sec:excitation}
Excitation distributions of fission fragment yields are estimated as follows.
First, total excitation energy (${\rm TXE}$) is calculated by
\begin{equation}
{\rm TXE}=E_{in} \frac{M_{\rm{U}}}{M_{\rm{U}}+M_{n}}+S_{n}+M_{f}-(M_{l}+M_{h})-{\rm TKE},
\end{equation}
where $E_{in}$ is the incident neutron energy, $S_{n}$ is neutron threshold of fissioning nucleus, and $M_{U}$, $M_{n}$, $M_{f}$, $M_{l}$, and $M_{h}$ are the mass of $^{235}$U, neutron, fissioning nucleus, light fragment, and heavy fragment, respectively.
To sort ${\rm TXE}$ into two fission fragments, we use an anisothermal model~\cite{Kawano2013}
\begin{equation}
    R_{T}=\frac{T_{l}}{T_{h}}
    =\sqrt{\frac{a_{h}(U_{h})U_{l}}{a_{l}(U_{l})U_{h}}},
    \label{eq:rt}
\end{equation}
where $U_{l,h}=E_{l,h}-\Delta$, $\Delta$ being a correction energy attributed from pairing correlations~\cite{Mengoni1994}.
The function $a_{l,h}(U_{l,h})$ is given by
\begin{equation}
a_{l,h}(U_{l,h})=f_{a} a^{*}_{l,h} \left(1+E_{sh:l,h}\frac{1-e^{-\gamma U_{l,h}}}{U_{l,h}}\right), 
\label{eq:levden}
\end{equation}
where $f_{a}$, $a^{*}$, $E_{sh:l,h}$, $\gamma$ are the scaling factor, the asymptotic level density parameter, the shell correction, and the shell damping factor, respectively.
The mean excitation energies $E_{l,h}$ are estimated numerically from Eq.~\eqref{eq:rt} with a condition of $E_{l}+E_{h}=\mathrm{TXE}$.
The excitation energy distributions are then estimated by~\cite{Okumura2021}
\begin{equation}
    F_{l,h}(E)=\frac{1}{\sqrt{2\pi}\sigma_{l,h}}\exp\left(-\frac{(E-E_{l,h})^{2}}{2\sigma_{l,h}^{2}}\right),
\end{equation}
where the width is given by
\begin{equation}
    \sigma_{l,h}
    =\frac{E_{l,h}}{\sqrt{E_{l}^{2}+E_{h}^{2}}}\sigma_{{\rm TKE}},
\end{equation} 
where $l$ and $h$ in the above equations indicate light and heavy fragments respectively. 
As a consequence, fission fragment of mass $A$, atomic number $Z$, and excitation energy $E$ is represented by
\begin{equation}
    Y(A,Z,E)=Y(A,Z)\times F_{A,Z}(E)
\end{equation}
In this work, $a^{*}, E_{sh},$ and $\gamma$ in Eq.~\eqref{eq:levden} are taken from Ref.~\cite{Mengoni1994}, while the scaling factor $f_{a}$ for fission fragments is sought by GP and GLS.
It is ideally desirable to seek $f_{a}$ for each fission fragments, however, we used the same $f_{a}$ for all fission fragments because experimental data are limited to determine it accurately.
\subsection{Odd-Even effects and Spin-Parity distributions}
From experimental data, one can observe that fission fragment yields of nuclei with even-proton or even-neutron are larger than those of neighboring nuclei with odd-proton and odd-neutron.
This effect is considered to be due to the pairing correlation; however its actual origin is not perfectly understood.
To consider the odd-even effect to fission fragment yields in this work, we multiply $Y(A,Z,E)$ by a factor of $f_{Z}$ $(f_{N})$ for even proton (neutron) number and divide for odd proton (neutron) number. 
After this correction, a renormalization is carried out to satisfy the condition:
\begin{equation}
\sum_{A,Z} \int Y(A, Z, E) \, dE=2.    
\end{equation}
\par
We assume equal distribution for odd and even parities.
The fission fragment yields including the spin-parity distribution are given by~\cite{GC}
\begin{equation}
Y(A,Z,E,J^{\pi})=\frac{1}{2}\frac{J+1/2}{2 f_{s}^{2}\sigma^{2}}
\exp \left(\frac{(J+1/2)^{2}}{2f_{s}^{2}\sigma^{2}(U)}\right) 
Y(A,Z,E).
\end{equation}
Here, the parameter $f_{s}$ adjusts the magnitude of the spin cutoff parameter $\sigma(U)$, which is determined by the parameter search of this study.
\subsection{Other input parameters}
Transmission coefficients of nucleons are calculated by the optical potentials of Koning-Delaroche~\cite{KandD2003}. 
$\alpha$, ${}^{3}$He, ${}^{3}$H and deuteron emissions from fission fragments are not considered in this work. 
Nuclear level schemes are taken from RIPL-3~\cite{Capote2009}.
For nuclear level densities, the Gilbert-Cameron method~\cite{GC} with Mengoni-Nakajima parameter~\cite{Mengoni1994} is adopted. 
For $\gamma$ strength functions, the enhanced generalized Lorentzian function~\cite{KandU} is used. Mass data are taken from the AME2020~\cite{Huang2021}.
\section{Seeking Parameters and Experimental data}
\label{sect:3}
In this section we explain the method to determine the parameters introduced in the previous section.
Let us define $\bm{x}\equiv(\{\delta Z_{p}\}, \{\delta \sigma_{p}\}, R_{T}, f_{Z}, f_{N}, f_{a}, f_{s})$.
The number of parameters is $113$ in total ($\delta Z_{p}$ and $\delta \sigma_{p}$ have $54$ parameters from $64\le A\le 117$, repsectively).
The parameter search is carried out by finding $\bm{x}$ that minimizes the objective function given by
\begin{equation}
O'(\bm{x})
=O(\bm{x})+G(\bm{x}).
\label{eq:objective}
\end{equation}
The first term of Eq.~\eqref{eq:objective} is identical to $\chi^{2}$-distribution and defined as
\begin{equation}
O(\bm{x})=\sum_{i}^{N_{\rm{exp}}}
\left|\frac{y_{exp}^{(i)}-y_{calc}^{(i)}(\bm{x})}{\delta y^{(i)}}\right|^{2},
\label{eq:objective1}
\end{equation}
where $N_{\rm{exp}}=1055$ is the number of experimental data.
The second term of Eq.~\eqref{eq:objective} is introduced to regulate the range of parameters and given by
\begin{equation}
    G(\bm{x})=\exp\left(\sum_{i}^{N_{par}}\frac{|x^{(i)}-x_{0}^{(i)}|^{2}}{2(\Delta x_{0}^{(i)})^{2}}\right).
    \label{eq:objective2}
\end{equation}
For $x_{0}$ and $\Delta x_{0}$ of $G(\bm{x})$ in Eq.~\eqref{eq:objective2}, we used
\begin{equation}
(x_{0}, \Delta x_{0})=
    \begin{cases}
        (0,\, 1.00)  \quad \mathrm{for} \quad \Delta Z_{p}\\
        (0,\, 0.25)  \quad \mathrm{for} \quad \Delta \sigma_{p}\\
        (1,\, 0.50)  \quad \mathrm{for} \quad R_{T}\\
        (1,\, 1.00)  \quad \mathrm{for} \quad f_{Z}, f_{N}\\
        (1,\, 0.50)  \quad \mathrm{for} \quad f_{a}\\
        (1,\, 4.00)  \quad \mathrm{for} \quad f_{s}\\
    \end{cases}
\end{equation}
\par
As already introduced, a Bayesian optimization with the Gaussian process is adopted as one of the parameter search methods, where the Gaussian process is adopted to predicts $O'(\bm{x})$ of Eq.~\eqref{eq:objective}.
The methodology is as follows.
We first calculate $O'(\bm{x})$ with several different $\bm{x}=\bm{x}_{inp}$ determined randomly and store the result as the initial input data ($\{\bm{x}_{inp}, O'_{inp}\}$).
Next, we predict the distribution of $O'(\bm{x})=O'_{pred}(\bm{x})$ within the Gaussian process using the input data.
From the predicted $O'_{pred}(\bm{x})$, we choose the most likely $\bm{x}=\bm{x}_{new}$ that gives the minimum $O'_{pred}(\bm{x})$ and it is added to the input data of $\{\bm{x}_{inp},O'_{inp}\}$.
%following input data points with an infinite dimensional Gaussian distribution.
Based on a Bayesian optimization, it is expected that one can find $\bm{x}$ that gives minimum $O(\bm{x})$ by iterating this process many times.
The formalism adopted in the present approach of GP is almost the same as~\cite{Watanabe2022}.
The difference from Ref.~\cite{Watanabe2022} is the objective function of Eq.~\eqref{eq:objective}.
\par
Another method to determine the most likely parameter set $\bm{x}$ is GLS.
Since outputs of the CCONE are generally non-linear against the input parameters, GLS is also carried out in an iterative way.
New parameter set is determined by adding $\delta \bm{x}$ to old parameter set $\bm{x}_{\mathrm{old}}$ with a regularization parameter $\mu$~$(0 < \mu \le 1)$ as
\begin{equation}
    \bm{x}_{\mathrm{new}}=\bm{x}_{\mathrm{old}}+\mu\, \delta \bm{x},
\end{equation}
where $\delta \bm{x}$ is determined from GLS:
\begin{equation}
\delta \bm{x}=
\mathrm{X}\mathrm{C}^{\rm{T}}
\left(
\mathrm{V}^{-1}(\bm{y}_{exp}-\bm{y}_{calc}(\bm{x}))
+\frac{(\bm{x}_{0}-\bm{x})}{2\Delta \bm{x}_{0}^{2}}G(\bm{x})
\right).
\end{equation}
The covariance matrix $\mathrm{X}$ is defined as
\begin{equation}
\mathrm{X}
=\left(
\mathrm{C}^{\rm{T}}\mathrm{V}^{-1}\mathrm{C}
+\frac{1}{2\Delta\bm{x}_{0}^{2}}G(\bm{x}) 
\right)^{-1},
\end{equation}
where $\mathrm{C}$ is the sensitivity matrix ($C_{ij}=\partial y_{calc,i}/\partial x_{j}$)and $\mathrm{V}$ $(\mathrm{V}_{ij}=1/(\delta y^{(i)})^{2}\delta_{ij})$ is the covariance matrix of experimental data.
In case of GLS, the initial $\bm{x}_{\mathrm{old}}$ are $\delta Z_{p}(A)=\delta \sigma_{p}(A)=0, R_{T}=1.0, f_{Z}=f_{N}=f_{a}=f_{s}=1.0$, while they are determined from GP in case of GP+GLS.
The regularization parameter $\mu$ is introduced so as to find the best parameter set smoothly and we give a different value of $\mu$ depending on the condition of $\delta \bm{x}$:
\begin{equation}
    \mu=
    \begin{cases}
        0.10 \quad (|\delta x^{(i)}/x_{\mathrm{old}}^{(i)}| <0.1)\\
        0.25 \quad (0.1 \le |\delta x^{(i)}/x_{\mathrm{old}}^{(i)}| <1)\\
        0.40 \quad (  1 \le |\delta x^{(i)}/x_{\mathrm{old}}^{(i)}| <10)\\
        0.20 \quad (\mathrm{other\,cases}).
    \end{cases}
\end{equation}
\par
Experimental data we used for the parameter search are listed in Table~\ref{tab:exp}.
We adopted independent and cumulative fission yield data by Rudstam et al.~\cite{1990ADNDT..45..239R}, Tipnis et al.~\cite{Tipnis1998}, prompt fission neutron yield data by Nishio et al,~\cite{Nishio1998}, Batenkov et al.~\cite{Batenkov2005}, Vorobyev et al.~\cite{Vorobyev2010}, Boldeman et al.~\cite{Boldeman1971}, Fraser et al.~\cite{Fraser1966}, Maslin et al~\cite{Maslin1967}, and G\"o\"ok~ et al.\cite{Gook2018}, averaged number of prompt $\gamma$ multiplicities for light and heavy fragments by Pleasonton et al.~\cite{Pleasonton1972}, decay heat data by Akiyama et al.~\cite{Akiyama}, Nguyen et al.~\cite{Nguyen1997}, Dickens et al.~\cite{Dickens}, and delayed neutron yield data by Keepin et al.~\cite{Keepin1957}.
The total number of experimental data point is $N_{\rm{exp}}=1055$.
\begin{table}
\centering
\caption{Experimental data used for parameter search.}
\begin{tabular}{|c|cc|}
\hline
Type & Author & Ref.\\
\hline
IFY & Rudstam et al.& \cite{1990ADNDT..45..239R}\\ 
    & Tipnis et al. & \cite{Tipnis1998} \\
\hline
CFY & Rudstam et al. & \cite{1990ADNDT..45..239R}\\
    & Tipnis et al. & \cite{Tipnis1998} \\
\hline
PFN & Nishio et al. & \cite{Nishio1998} \\
    & Batankov et al.& \cite{Batenkov2005} \\
    & Vorobyev et al. & \cite{Vorobyev2010} \\
    & Boldeman et al. & \cite{Boldeman1971} \\
    & Fraser et al. & \cite{Fraser1966} \\
    & Maslin et al. & \cite{Maslin1967} \\
    & G\"o\"ok et al. & \cite{Gook2018} \\
\hline
Average number of PFG & Pleasonton et al. & \cite{Pleasonton1972} \\
\hline
Decay Heats after & Akiyama et al. & \cite{Akiyama} \\
neutron irradiation  & Nguyen et al. & \cite{Nguyen1997} \\
                    & Dickens et al. & \cite{Dickens} \\
\hline
Delayed Neutrons after & Keepin et al. & \cite{Keepin1957}\\
neutron irradiation    &        &   \\
\hline
Total Delayed Neutron Yield & Keepin et al. & \cite{Keepin1957}\\
\hline
\end{tabular}
\label{tab:exp}
\end{table}
\par
We have $54$ free parameters for $Z_{p}(A)$ and $\sigma_{p}(A)$, respectively.
If either $\delta Z_{p}(A)$ or $\delta \sigma_{p}(A)$ for a certain $A$ is insensitive to the experimental data, we set $\delta Z_{p}(A)=0$ or $\delta \sigma_{p}(A)=0$ and exclude from the parameter fitting procedure.
This case occurs especially for the far-wing part ($A\le70$ and $A\ge166$) because of lack of experimental data in these regions.
\section{Result}
\label{sect:4}
A test calculation is carried out for a thermal neutron-induced fission on $^{235}$U.
The parameter search introduced in Sect.~\ref{sect:3} has to be carried out to obtain independent fission fragment yields and fission observables reproducing experimental data.
Before going to a parameter search for full set of $\bm{x}=(\{\delta Z_{p}\},\{\delta \sigma_{p}\}, R_{T}, f_{Z}, f_{N}, f_{a}, f_{s})$, we checked the performance of the parameter search method described in Sect.~\ref{sect:3} by reducing the number of parameters.
Assuming that $\delta Z_{p}(A)$ and $\delta \sigma_{p}(A)$ have a smooth dependence on $A$, we divide them into $5$ parts, namely, $64 \le A \le 74$, $75 \le A \le 85$, $86 \le A \le 96$, $\le 97 \le A \le 107$, and $108 \le A \le 117$.
Finally we discuss the result of full parameter set of of $\bm{x}$.
\subsection{Test of parameter search with a reduced number of parameters}
\label{sect:5}
\begin{figure}[hbtp]
\centering
\includegraphics[width=0.98\linewidth, bb=0 0 620 490]{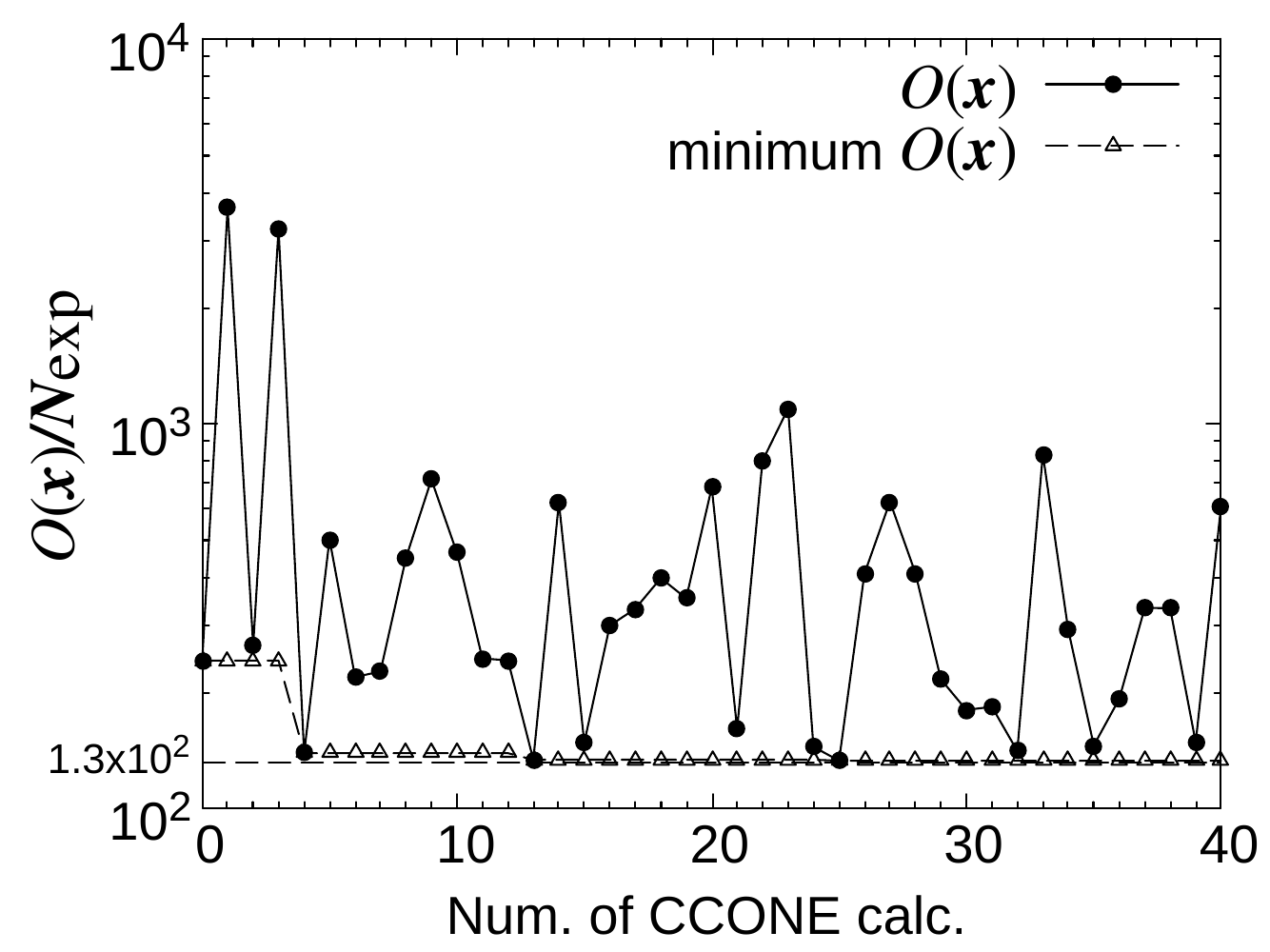}
\caption{Objective function normalized to $N_{\rm{exp}}$ calculated by GP (circle). The $x$-axis represents the number of CCONE calculation. The minimum $O(\bm{x})/N_{\rm{exp}}$ in the past calculations is indicated by the triangle.}
\label{fig:GP5}
\end{figure}

In this section, the result of the parameter search method for $\bm{x}=(\{\delta Z_{p}\}, \{\delta \sigma_{p}\},R_{T}, f_{Z}, f_{N}, f_{a}, f_{s})$ is discussed, where $\delta Z_{p}(A)$ and $\delta \sigma_{p}(A)$ ($64\le A \le 117$) are divided into $5$ parts.
Thus, the number of parameters to be determined is $15$ in total.
We first carry out the parameter search by GP.
The result of $O(\bm{x})$ is shown in Fig.~\ref{fig:GP5}, where the $x$-axis is the number of CCONE calculation, which is identical to the iteration time of GP.
The minimum $O(\bm{x})$ that has been obtained in the past calculations is indicated by the triangle.
The objective function oscillates around $O(\bm{x})=200$ and does not get converged even after $40$ times of the calculations.
We confirmed that $O(\bm{x})$ did not converge even if we increase the number of iteration up to $200$.
Analyzing the history of $\bm{x}$ in this parameter search, GP tries various kinds of $\bm{x}$ in the $15$-dimension parameter space.
In general, GP requires many times of calculations until one gets a convergent result especially for a multi-dimensional case like this.
However, we notice that a small $O(\bm{x}) (\simeq 130)$ was already found after a few times of the CCONE calculations.
From this result, GP may provide one of the candidates of $\bm{x}$.
\par
\begin{figure}
\centering
\includegraphics[width=0.3\linewidth, bb=400 100 550 440]{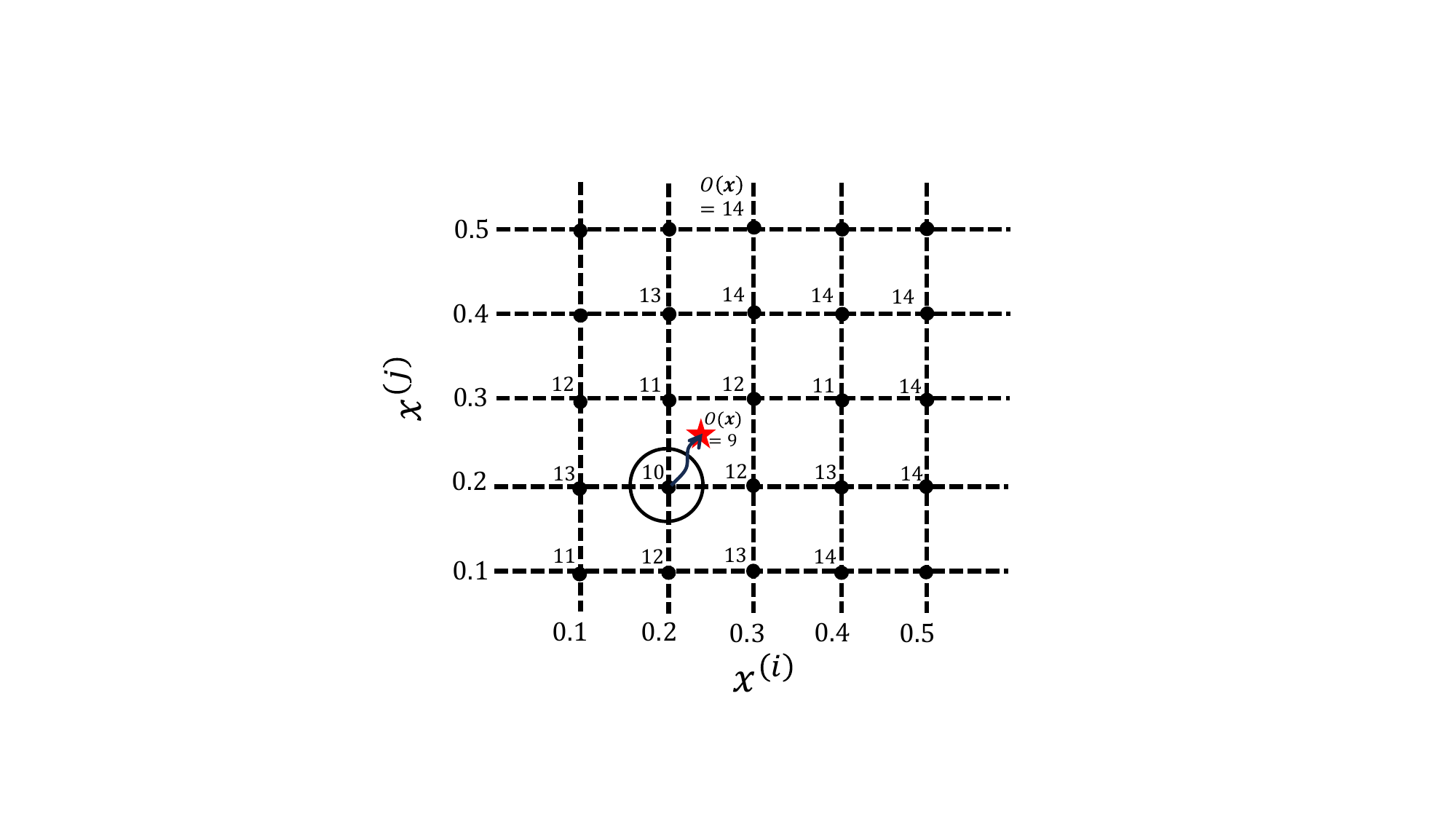}
\caption{Schematic figure explaining parameter search by a combination of GP and GLS used in this work. 
First of all, a multi-dimensional parameter space is discretized by an arbitrary grid size ($0.1$ in this figure). Then, we seek a minimum $O(\bm{x})$ on the grid points by GP.
The result is shown by the open circle ($\bm{x}_{\mathrm{GP}}=[x^{(i)},x^{(j)}]=[0.2,0.2])$.
Then, a more realistic minimum $O(\bm{x})$ shown by a star symbol is sought by GLS setting $\bm{x}_{\mathrm{GP}}$ as an initial parameter set.}
\label{fig:Grid}
\end{figure}

On the other hand, a parameter search by GLS has a shortcoming that it is often stack into to a local minimum of $O(\bm{x})$ unless an appropriate initial parameter set is chosen.
To avoid this problem, one needs to change the initial parameter set in several time and check if a consistent result is obtained.
However, this procedure also costs a lot of calculations.
To overcome this problem, we come up with an idea to combine GP and GLS.
A schematic figure is shown in Fig.~\ref{fig:Grid} to illustrate the new approach.
The procedure is as follows.
We first consider a discrete mesh of the parameter space (a two-dimensional mesh is drawn in Fig.~\ref{fig:Grid}; however, actual calculations are carried out in multi-dimensional mesh).
We next seek a minimum $O(\bm{x})$ on the grids of the mesh within GP, which is shown by the open circle ($\bm{x}_{\mathrm{GP}}=[x^{(i)},x^{(j)}]=[0.2,0.2]$) in Fig.~\ref{fig:Grid}.
A real minimum $O(\bm{x})$ is expected to be around this grid point and is sought by GLS setting $\bm{x}_{\mathrm{GP}}$ as an initial parameter set.
If this process works well, an expected result would be obtained.
In what follows, we call the combination of GP and GLS as GP+GLS.
\par
\begin{figure}[hbtp]
\centering
\includegraphics[width=0.98\linewidth, bb=0 0 620 490]{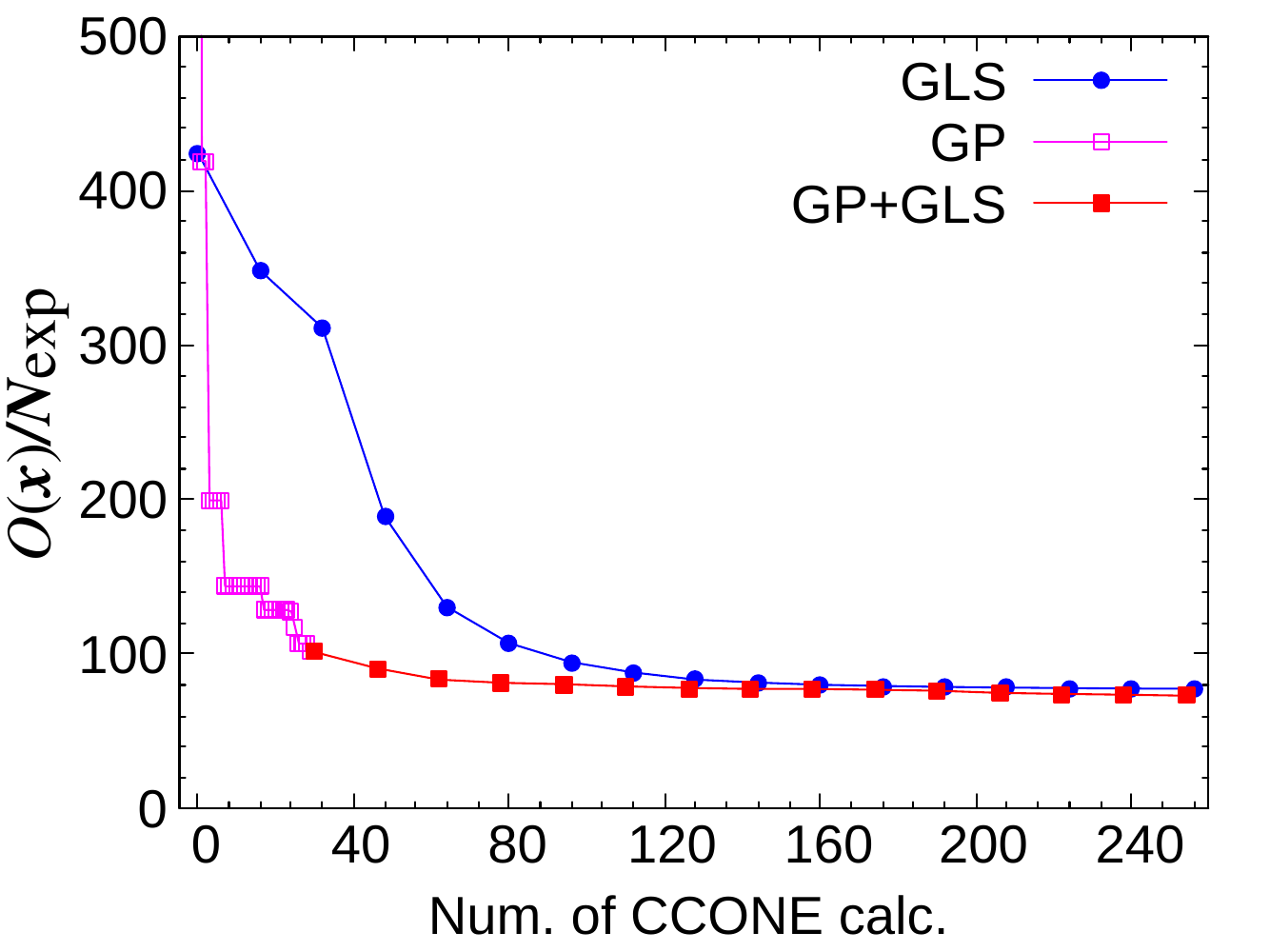}
\caption{Objective function normalized to $N_{\rm{exp}}$ as a function of the number of the CCONE calculation. GP+GLS (solid square) is started from the calculation using the parameter set obtained by GP.}
\label{fig:GPGLS5}
\end{figure}
\begin{figure}[hbtp]
\centering
\includegraphics[width=0.98\linewidth, bb=0 0 620 490]{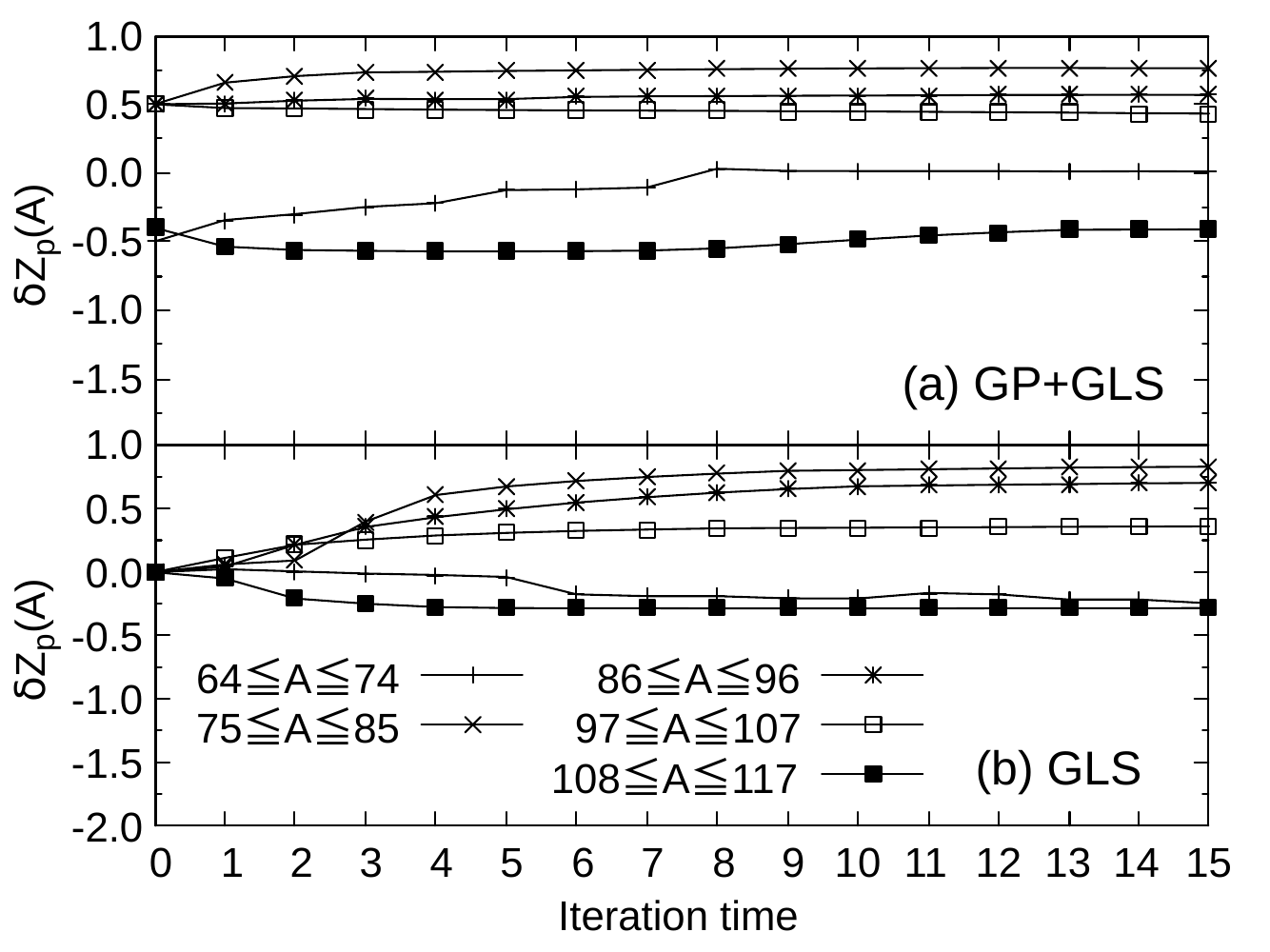}
\caption{$5$ $Z_{p}(A)$ parameters at each iteration step of the least square fitting for (a) GP+GLS and (b) GLS.}
\label{fig:par5}
\end{figure}

Figure~\ref{fig:GPGLS5} shows the objective function obtained with the parameter search methods of GLS and GP+GLS.
The GP+GLS calculation is launched by using $\bm{x}_{\mathrm{GP}}$ as explained above.
We can see that the objective function of $O(\bm{x})/N_{\rm{exp}}$ of GP+GLS converges much faster than that of GLS indicated by the closed circle.
Furthermore, GP+GLS gives a lower $O(\bm{x})/N_{\rm{exp}}$ ($\simeq 72$) than that of GP ($\simeq 77$), proving that GP+GLS is more effective than GLS.
We depicted $5$ $\delta Z_{p}$ parameters, which are sensitive to fission fragment yields, at each iteration time of GLS in Fig.~\ref{fig:par5}.
Note that the initial values of GP+GLS (iteration time $0$) is identical to the result of GP, so that the symbols shown in Fig.~\ref{fig:par5} is the result after Bayesian optimization with the Gaussian process.
As expected, most of $\delta Z_{p}$ parameters of GP+GLS does not change significantly at each iteration step.
This indicates GP could find $\delta Z_{p}$ close to reasonable values.
Only $\delta Z_{p}$ for $64\le A\le 74$ for GP+GLS shifts from $\delta Z_{p}\sim-0.5$ at the beginning to $\delta Z_{p}\sim 0.0$ at the end.
This parameter ranges to the far-wing region of fission fragment yields and does not affect the calculated fission observables, namely its sensitivity to the experimental data is limited.
In fact, $\delta Z_{p}$ for $64\le A\le 74$ as well as $\delta \sigma_{p}$ has a sensitivity only for $12$ PFN experimental data, while other parameters usually have a sensitivity for more than hundred experimental data.
It is difficult to determine such a parameter uniquely within the limited number of iteration time by GP.
However, this issue is remedied by GLS and the parameters are accommodated to the experimental data more precisely.
For the case of GLS, $\delta Z_{p}$ parameters show meaningful changes at first few iterations.
After $15$ time iteration, the results of GLS become closer to those of GP+GLS, but have some differences leading to the larger objective function as seen in Fig~\ref{fig:GPGLS5}.
The symmetric part ($108 \le A \le 117$) show negative $\delta Z_{p}$, which is consistent to Wahl's evaluation shown in Fig.~\ref{fig:Zp}, while the far-wing part ($64 \le A \le 74$) is close to zero (GP+GLS) or negative (GP), which contradicts to Wahl's evaluation.
\par
\begin{figure}[hbtp]
\centering
\includegraphics[width=0.98\linewidth, bb=90 50 540 450]{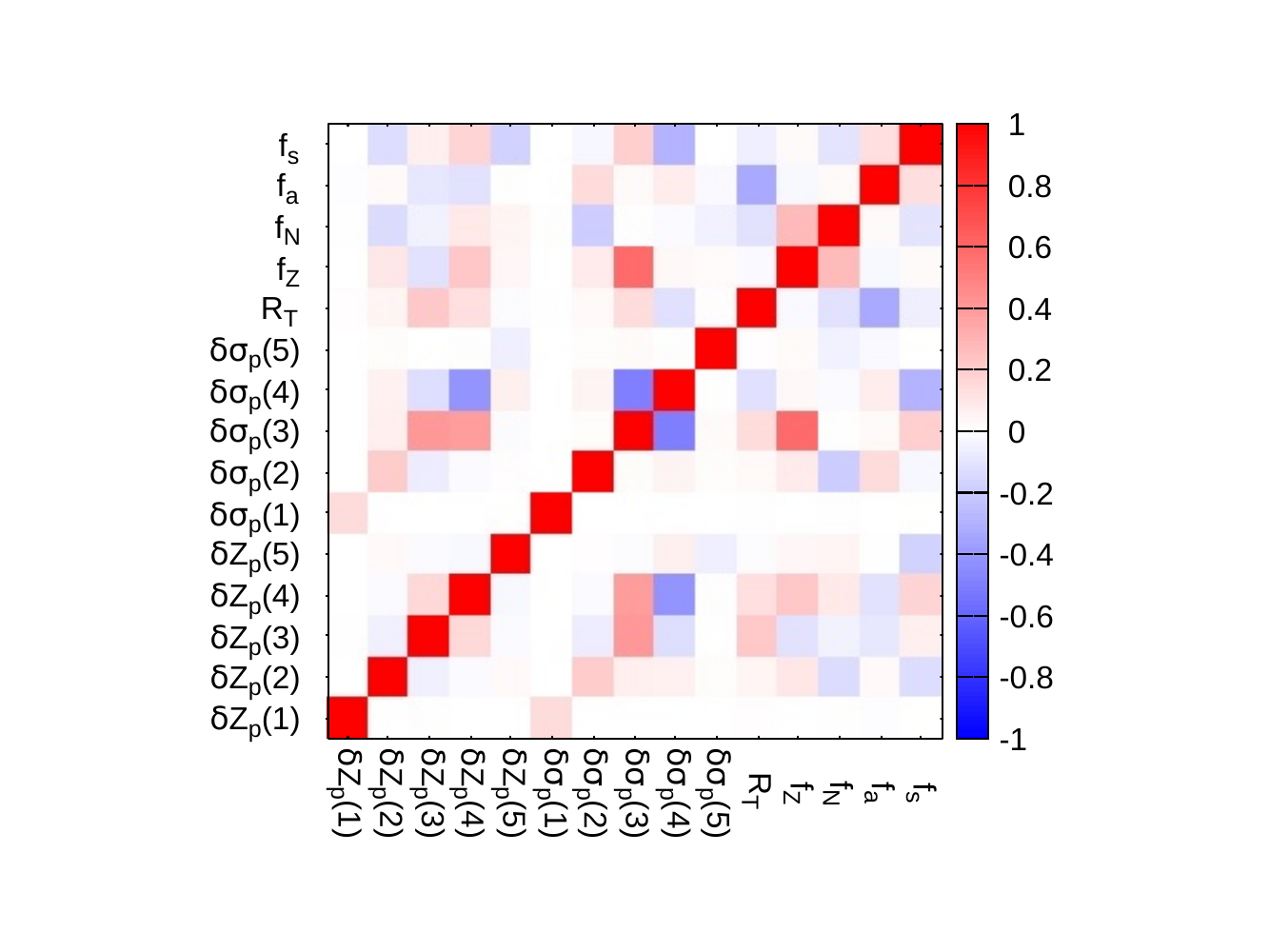}
\caption{Correlation matrix of the parameters of $\bm{x}=(\{\delta Z_{p}\}, \{\delta \sigma_{p}\},R_{T}, f_{Z}, f_{N}, f_{a}, f_{s})$.}
\label{fig:Par5Cov}
\end{figure}

We plot the correlation matrix of the $15$ parameters in Fig~\ref{fig:Par5Cov}.
Here, the correlation coefficients $\mathrm{COR}(X,Y)$ are defined as $\mathrm{COR}(X,Y)=\frac{\mathrm{COV}(X,Y)}{\sigma(X)\sigma(Y)}$, where $\mathrm{COV}(X,Y)$ are between two variables of $X$ and $Y$ and $\sigma(X)$ are the standard deviations of $X$.
The symbols of $\delta Z_{p}(i)$ and $\delta \sigma_{p}(i)$ $(i=1,\cdots, 5$) in Fig.~\ref{fig:Par5Cov} are the corresponding parameters in the mass regions of $(1)64 \le A \le 74$, $(2)75 \le A \le 85$, $(3)86 \le A \le 96$, $(4)\le 97 \le A \le 107$, and $(5)108 \le A \le 117$.
We should note that $\delta Z_{p}$ and $\sigma_{p}$ in the same mass region have relatively high correlations as is obvious from Eq.~\eqref{eq:charge}.
Another remarkable point is that $\sigma_{p}(3)$ and $\sigma_{p}(4)$ show a relatively strong correlation of about $\mathrm{COR}(X,Y)=-0.54$ despite that they belong to different mass regions.
Actually, neighboring mass regions may have a correlation because some fission observables accompany particle emissions, e.g. PFN, decay heats, and delayed neutron yields.
In addition, the mass regions of (3) and (4) correspond to $86 \le A \le 107$ where fission fragment yields are high, and are sensitive to fission observables.
For these reasons, $\sigma_{p}(3)$ and $\sigma_{p}(4)$ have a relatively high correlation.
However, most pairs of parameters have small correlations and this facts enables us to determine the parameters relatively smoothly.
\par
\begin{figure}[hbtp]
\centering
\includegraphics[width=0.98\linewidth, bb=0 0 620 490]{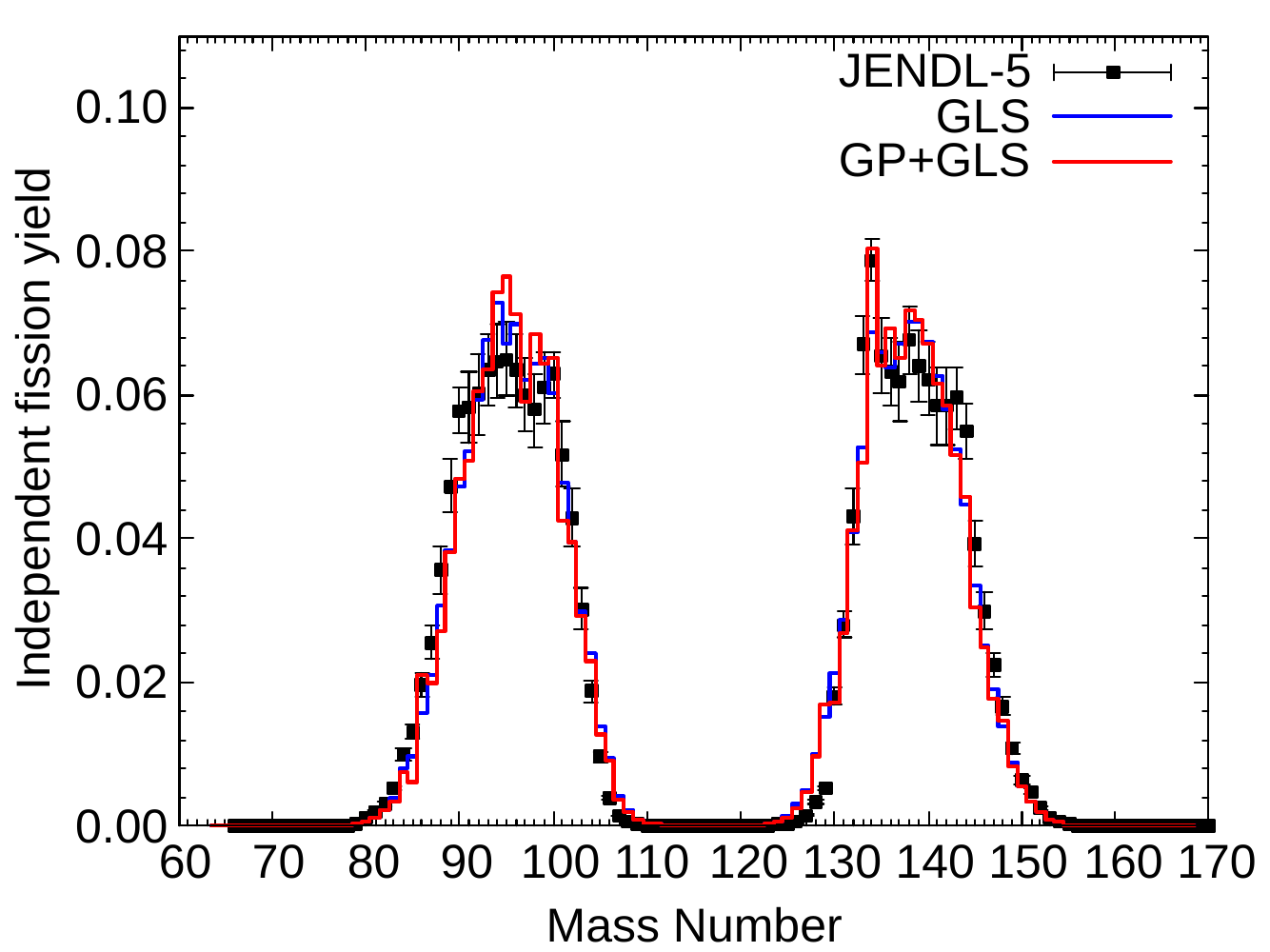}
\caption{Independent fission mass yields calculated with the parameter search methods of GP and GP+GLS. The evaluated data of JENDL-5 are also shown.}
\label{fig:ind5}
\end{figure}
\begin{figure*}[hbtp]
\centering
\includegraphics[width=1.0\linewidth, bb=0 0 650 300]{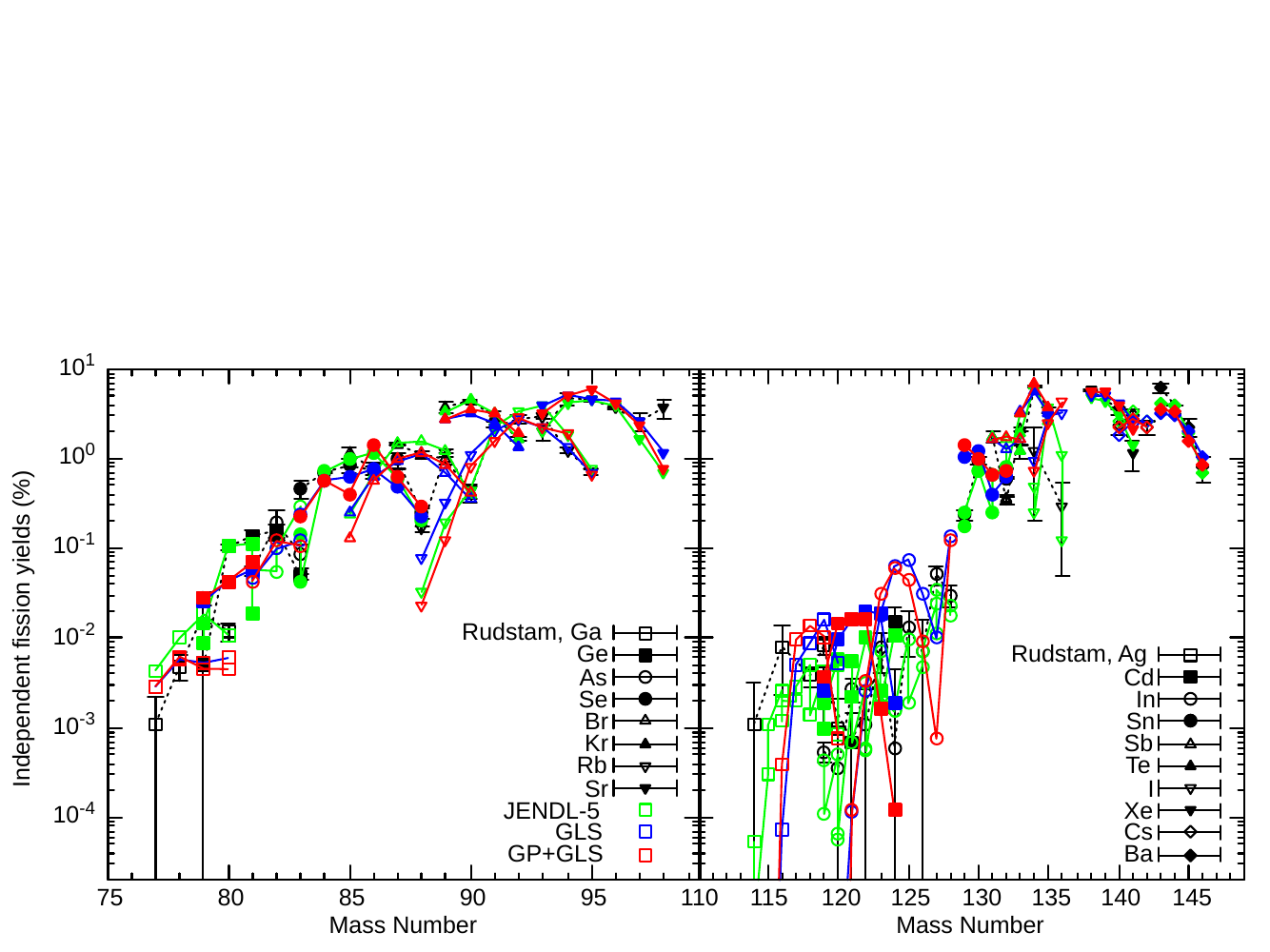}
\caption{Independent fission fragment yields.}
\label{fig:rud5}
\end{figure*}

Next, we compare the calculated fission observables with experimental and evaluated data.
The result of independent fission fragment mass yields are shown in Fig~\ref{fig:ind5}, where the evaluated data of JENDL-5~\cite{JENDL-5} are shown together.
The results of GP, GP+GLS, and JENDL-5 show a good agreement to each other.
However, both GP and GLS provide higher yields than JENDL-5 in the mass region from $92\le A \le 98$.
This difference must be compensated because independent fission fragment mass yields are normalized to $2$.
Actually, the fragment distributions of GP and GLS are slightly narrower than JENDL-5 .
Figure~\ref{fig:rud5} illustrates the result of IFY of fragments.
The experimental data of Rudstam~\cite{1990ADNDT..45..239R} are also shown for comparison.
Despite that we reduced the number of adjustable parameters for $\delta Z_{p}(A)$ and $\delta \sigma_{p}(A)$, both the calculated result of GP and GP+GLS reproduce the experimental data reasonably.
%This indicates that IFY do not strongly depend on the fine structure of charge distribution.
%
\par
\begin{figure}[hbtp]
\centering
\includegraphics[width=0.98\linewidth, bb=0 0 700 550]{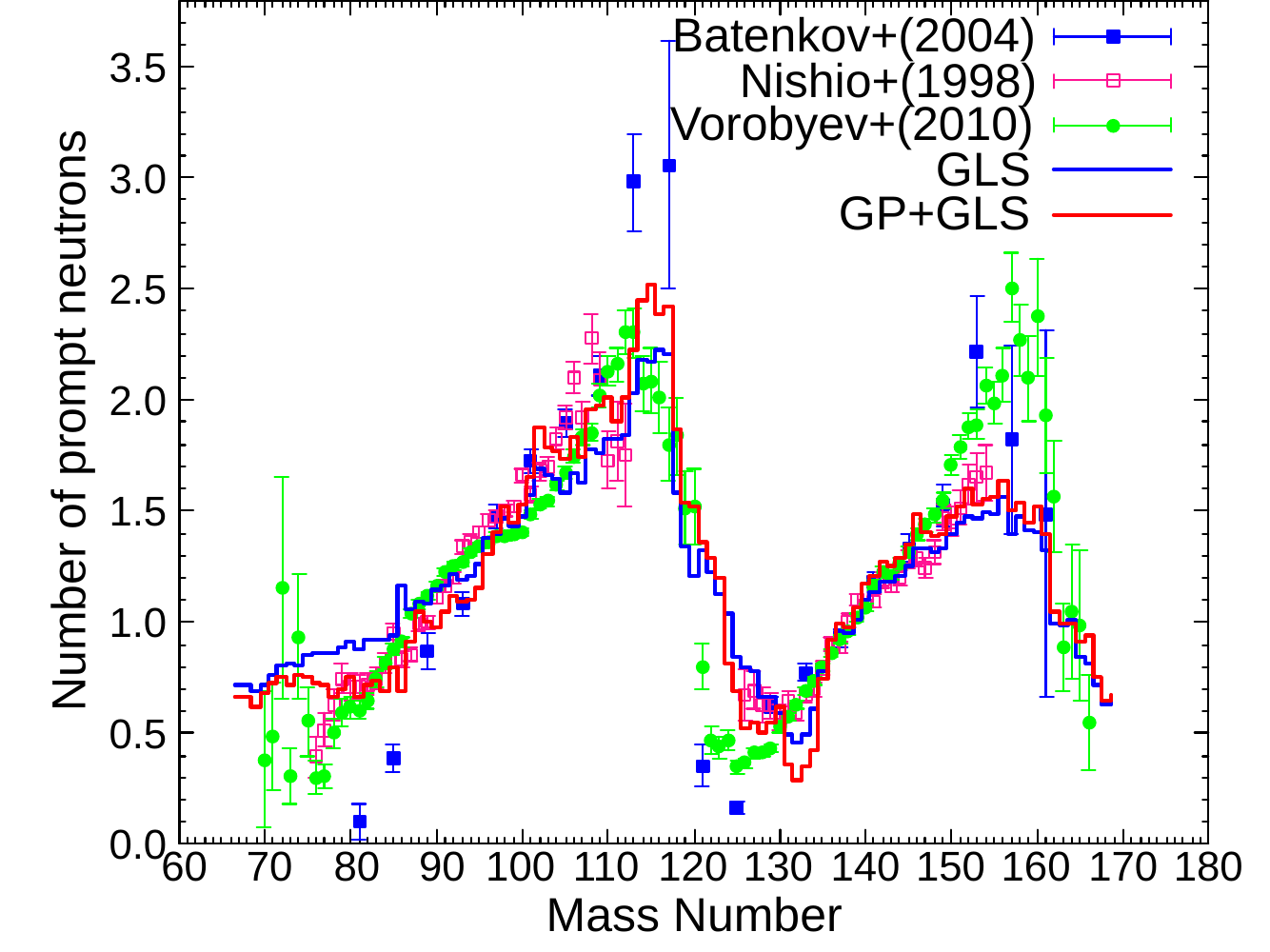}
\caption{Number of prompt neutron multiplicity as a function of fission fragment mass in case of $5$ parameters for $\delta Z_{p}$ and $\delta \sigma_{p}$.}
\label{fig:pnapga5}
\end{figure}
\begin{figure}[hbtp]
\centering
\includegraphics[width=0.98\linewidth, bb=0 0 350 470]{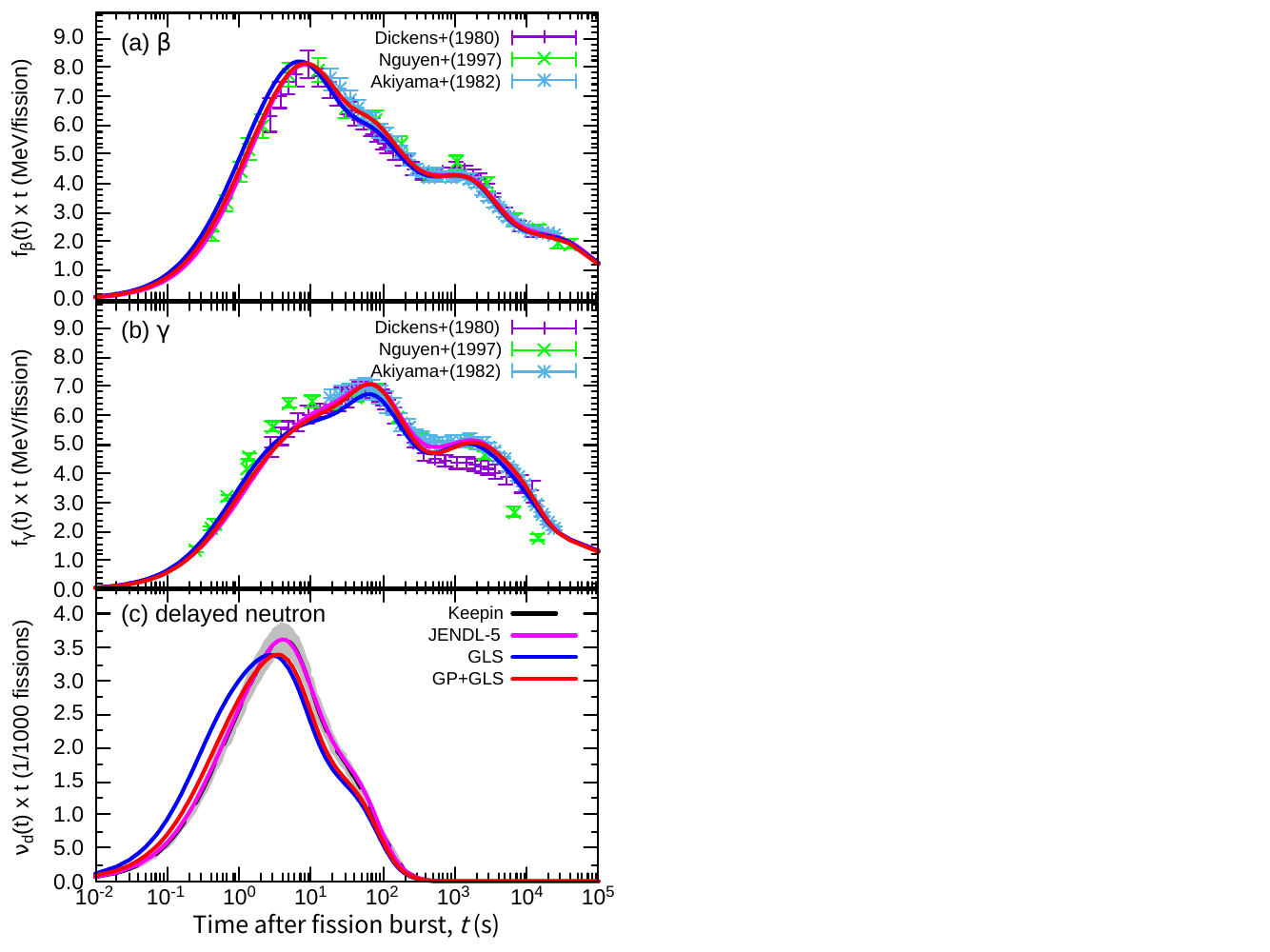}
\caption{(a)$\beta$-ray and (b)$\gamma$-ray decay heats and (c) delayed neutron yields as a function of time after fission burst (instant neutron radiation). The evaluated result of JENDL-5 (magenta) is also shown together.}
\label{fig:oyak5}
\end{figure}

Figure~\ref{fig:pnapga5} compares the number of prompt neutron multiplicity as a function of fission fragment mass.
The experimental data exhibit the so-called saw-tooth structure, namely the numbers of prompt-neutron increase from $A=70$ to $110$ in light fragments, and sudden decrease when jumping to heavy fragments, but again increase from $A=120$ to $160$.
Although the calculated result successfully reproduces the experimental data, especially those of Nishio et al.~\cite{Nishio1998}.
In Fig.~\ref{fig:oyak5}, the results of (a) $\beta$-ray and (b) $\gamma$-ray decay heats and (c) delayed neutron yields as a function of time after an instant neutron irradiation are shown.
The decay heats of $\beta$- and $\gamma$-rays are reproduced well.
On the other hand, the result of delayed neutron yields is not reproduced well both for GLS and GP+GLS.
Delayed neutron yields are rather sensitive to the charge distribution~\cite{Minato2017} and we confirmed that the $5$ parameters of $\delta Z_{p}$ and $\sigma_{p}$ are not enough to reproduce the experimental data.
\subsection{Full parameter set for $\delta Z_{p}(A)$ and $\delta \sigma_{p}(A)$}
\label{sect:full}
\begin{figure}[hbtp]
\centering
\includegraphics[width=0.98\linewidth, bb=0 0 620 490]{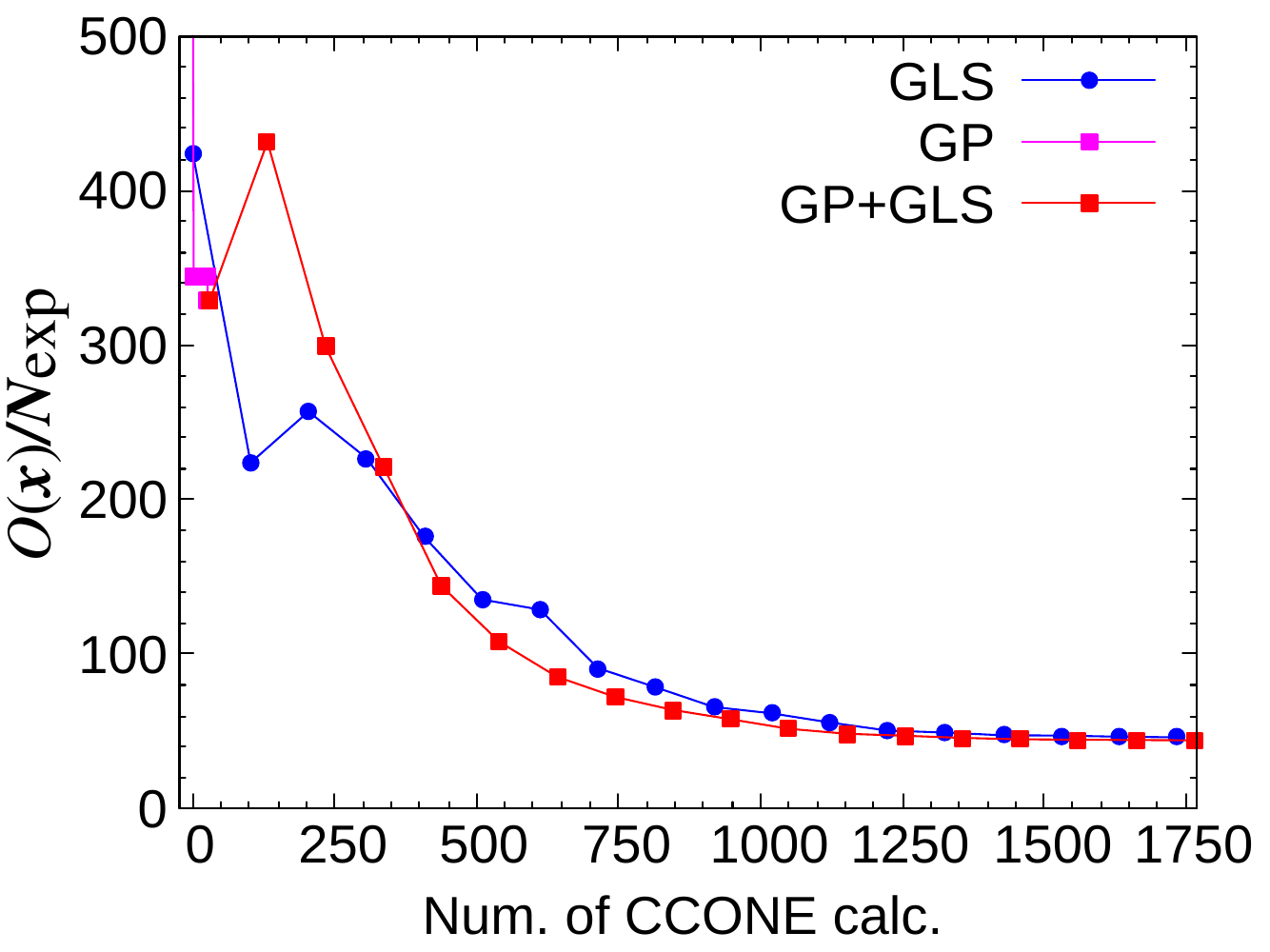}
\caption{Objective function textcolor{red}{normalized to $N_{\rm{exp}}$} and the required number of CCONE calculation.}
\label{fig:GPGLS54}
\end{figure}

\begin{figure}
    \centering
    \includegraphics[width=0.98\linewidth, bb=0 0 600 500]{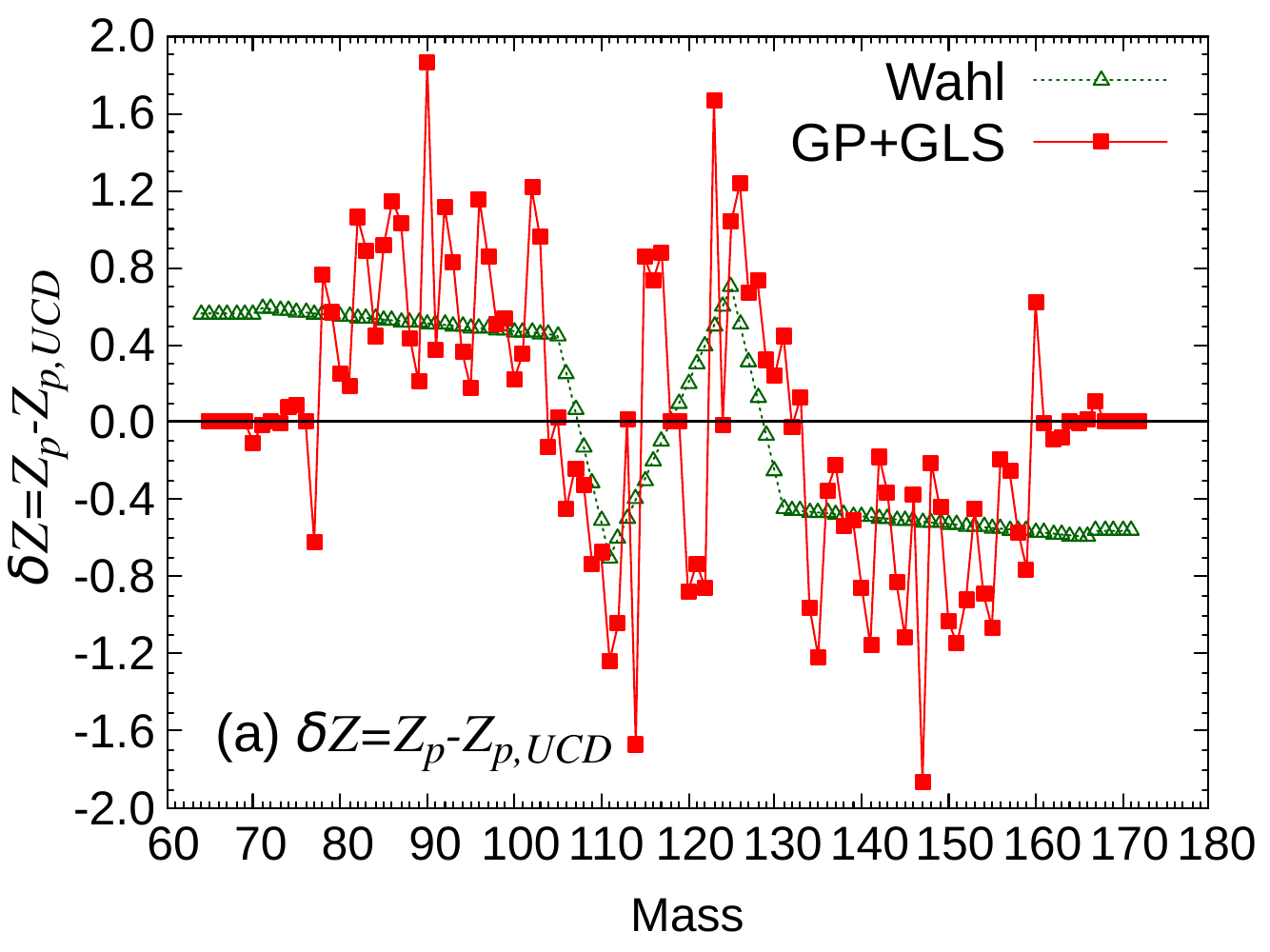}
    \includegraphics[width=0.98\linewidth, bb=0 0 600 500]{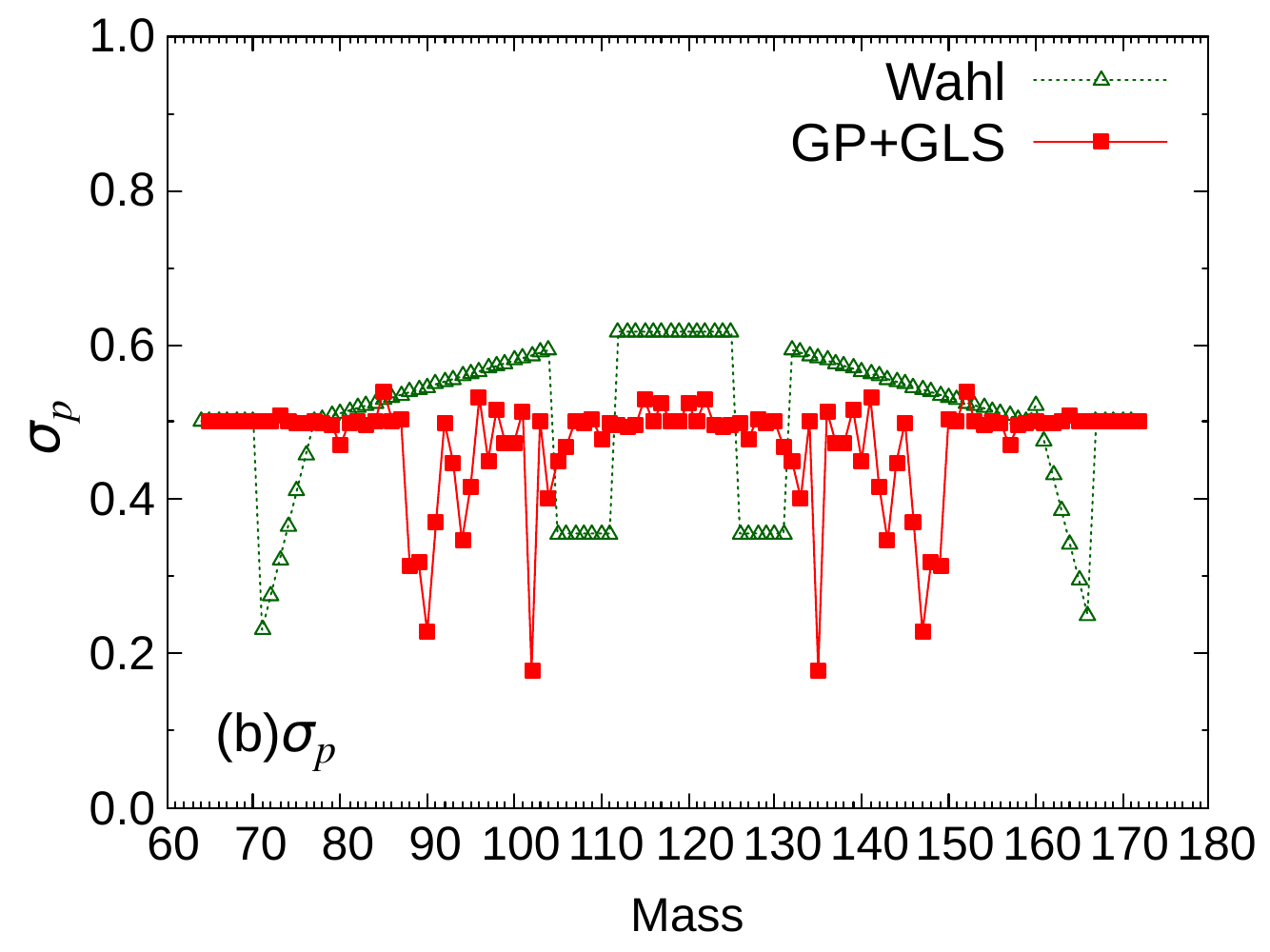}
    \caption{(a) Difference between unchanged charge distribution and evaluated value of the most probable proton number $Z_{p}$. 
    The calculated results of GP+GLS are compared with the evaluated data of Wahl~\cite{Wahl2002}. 
    (b) Width parameters of charge distributions.}
    \label{fig:Zp54}
\end{figure}

Now we demonstrate the result of full $54$ parameter set for for $\delta Z_{p}(A)$ and $\delta \sigma_{p}(A)$, namely $113$ parameters in total.
We first discuss the result of objective function normalized to $N_{\rm{exp}}$ at each iteration step.
The result is shown in Fig.~\ref{fig:GPGLS54}.
As in the case of $5$ parameters for $\delta Z_{p}(A)$ and $\delta \sigma_{p}(A)$, GP+GLS starts with the parameter set obtained after $30$ iterations of GP.
The objective function of GP+GLS decreases more quickly than that of GP and converges to $O(\bm{x})/N_{\rm{exp}}\sim 44$, while the objective function of GLS converges to $O(\bm{x})/N_{\rm{exp}}\sim 46$.
Thus, it can be concluded that GP+GLS can find a better result than GLS even when the number of free parameter increases.
Hereafter, we only show the results of GP+GLS regarding that its performance is better than those of GLS.
We compare $\delta Z_{p}(A)$ and $\sigma_{p}(A)$ obtained by the parameter search and Wahl's evaluation in Fig.~\ref{fig:Zp54}.
Note that a part of the far-wing regions, which are $64\le A \le 69$ and $169 \le A \le 174$, is excluded from the parameter search target because they are insensitive to any of the experimental data listed in Table~\ref{tab:exp} and cannot be adjusted by the GLS.
The calculated result has a staggering distribution, but the average behavior shows a similar trend to Wahl's evaluation.
It is interesting that the saw-tooth structures evaluated by Wahl around $A=110$ and $126$ are also confirmed by the present parameter search method as well.
On the other hand, $Z_{p}(A)$ at $70 \le A \le 75 (163 \le A \le 168)$ are close to zero for the present result, showing a difference from the Wahl's evaluation.
As for $\sigma_{p}(A)$, the present result shows a completely different result from the Wahl's evaluation.
For $70 \le A \le 86$, $\sigma_{p}(A)$ are almost equal to $0.5$ and they become $<0.5$ with some staggering behavior for $86 \le A \le 110$.
These structure may arise from the shell structure of nuclei at fissioning point.
A increase as estimated by Wahl from $A=75$ to $104$ and a plateau around the symmetric region are not obtained for the present model.
%We can see also that differences between GLS and GP+GLS are not large in this case.
%
%%
%\par
%%
%Figure~\ref{fig:Par54Cov} shows the correlation matrix of the parameters obtained by GP+GLS.
%Since the number of parameters is large, the sectors of parameter type, namely, $\delta Z_{p}(A)$, $\delta \sigma_{p}(A)$, and ($R_{T},f_{Z}, f_{N}, a, f$) are shown.
%As seen in Fig.~\ref{fig:Par5Cov}, correlations are small between different parameters.
%%
%\begin{figure}[hbtp]
%\centering
%\includegraphics[width=0.8\linewidth, bb=70 40 530 450]{Par54Cov.pdf}
%\caption{Correlation matrix of the parameters of $\bm{x}=(\{\delta Z_{p}\}, \{\delta \sigma_{p}\},R_{T}, f_{Z}, f_{N}, f_{a}, f_{s})$.}
%\label{fig:Par54Cov}
%\end{figure}
%
\par
\begin{figure}[hbtp]
\centering
\includegraphics[width=0.98\linewidth, bb=0 0 620 480]{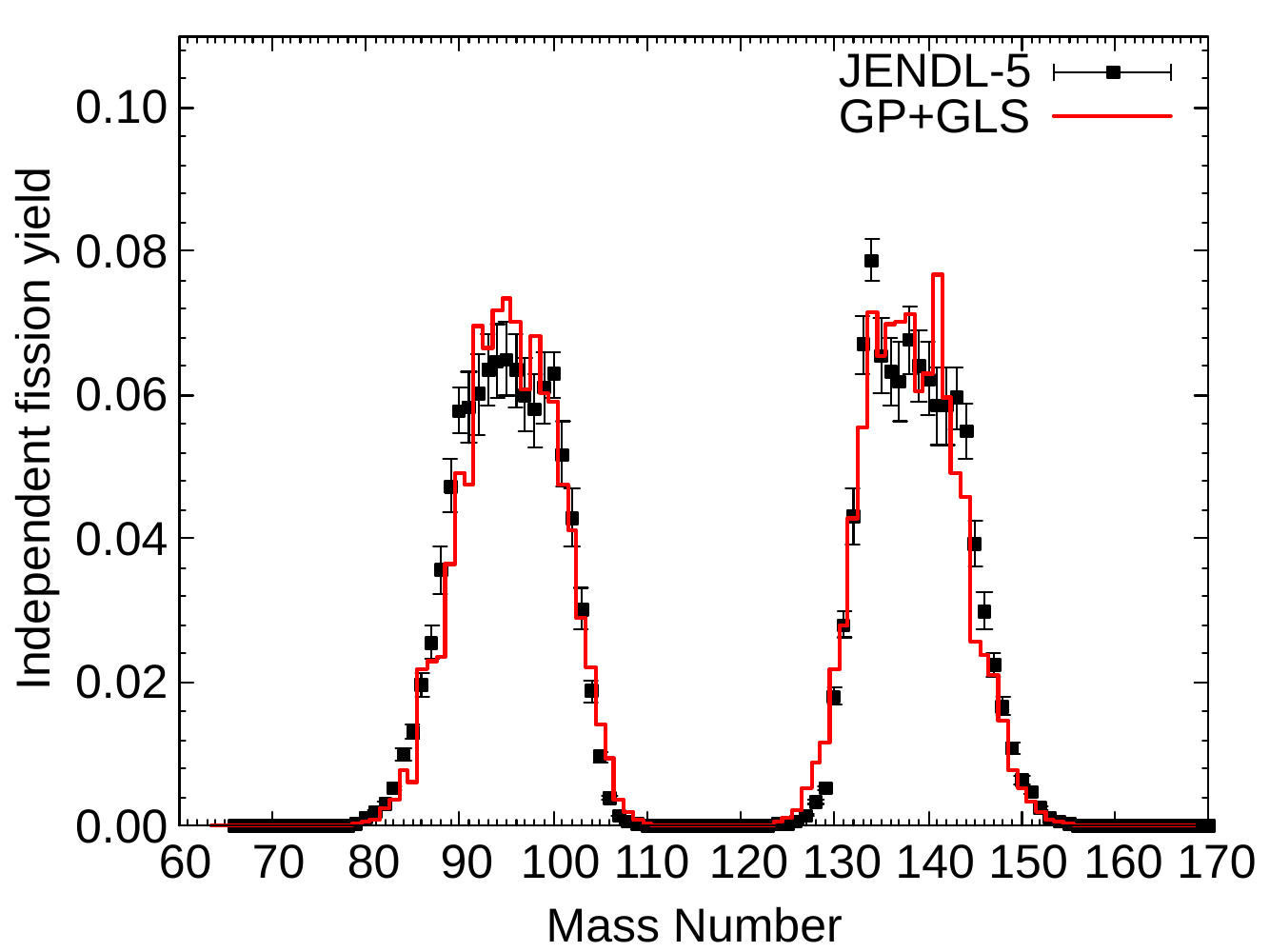}
\caption{Independent fission mass yields for full parameter set for $\delta Z_{p}(A)$ and $\delta \sigma_{p}(A)$.}
\label{fig:ind54}
\end{figure}
\begin{figure*}[hbtp]
\centering
\includegraphics[width=1.0\linewidth, bb=0 0 650 300]{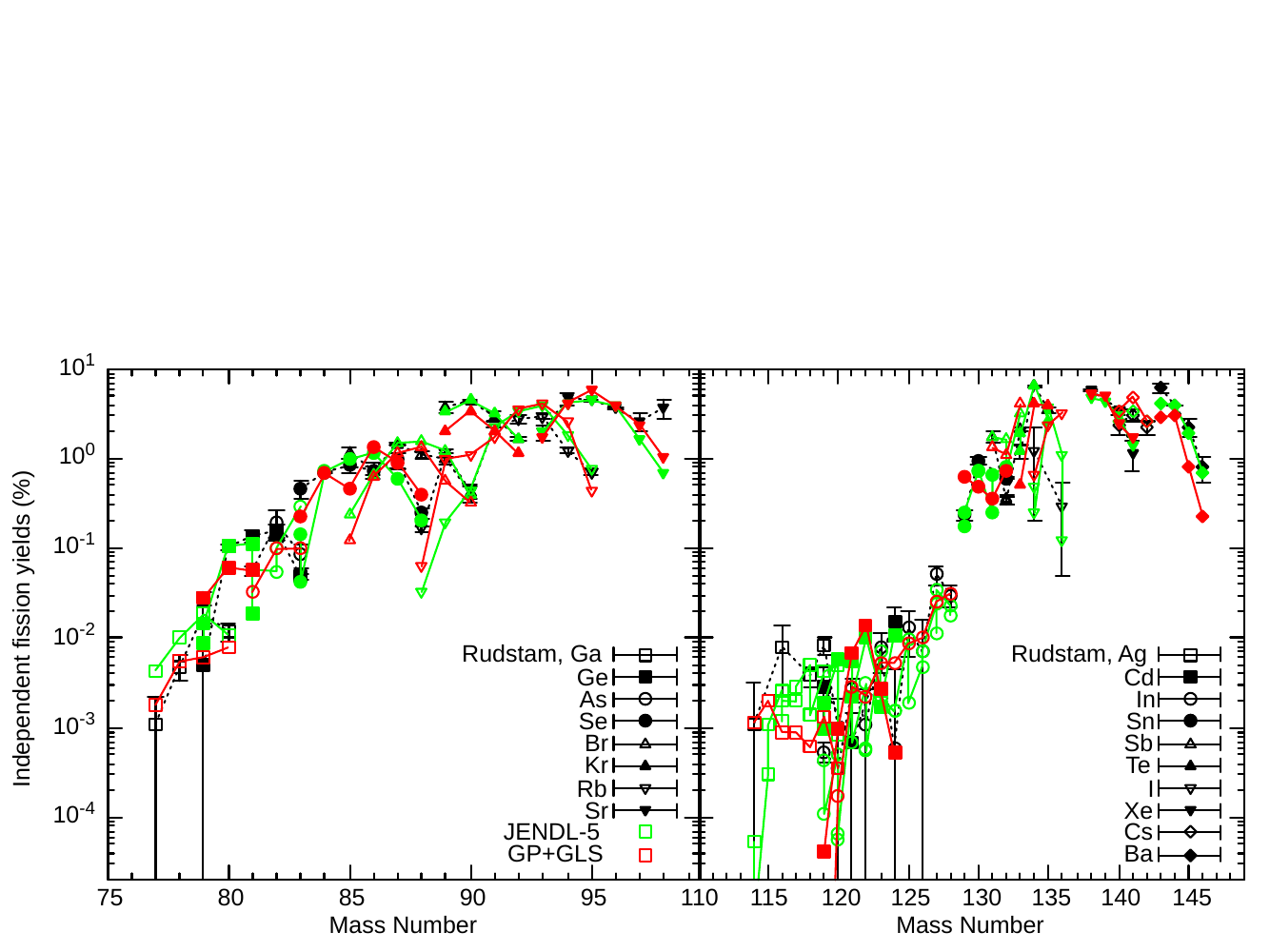}
\caption{Same as Fig.~\ref{fig:ind5}, but for full parameter set for $\delta Z_{p}(A)$ and $\delta \sigma_{p}(A)$.}
\label{fig:rud54}
\end{figure*}

Figure~\ref{fig:ind54} shows the independent fission mass yields.
One of the remarks of the GP+GLS result is that it shows a peak at $A=141$, which was not observed for JENDL-5.
The peak is mainly due to IFY of $^{141}$Cs, which is 4.9\% for GP+GLS and 3.3\% for JENDL-5.
From the analysis, the pre-neutron fission fragment yield of $^{142}$Cs is about 4.9\% in the present calculation (about 3.4\% for $^{141}$Cs).
The high pre-neutron fission fragment yield of $^{142}$Cs is attributed from $Z_{p}(A)=55.044$ after the parameter fitting.
As a result, the feeding to $^{141}$Cs by 1-neutron emission works effectively to make the peak.
We also see that the independent fission mass yields for light fragments for $90\le A \le 95$ are slightly larger than those for JENDL-5.
This was already found in Fig.~\ref{fig:ind5} and was also confirmed in other work of Ref.~\cite{Okumura2018} that adopts the Hauser-Feshbach statistical model to calculate independent fission mass yields from pre-neutron fission yields.
%Since our study considers neutron emissions in a more accurate way than JENDL-5, this may indicate that the evaluated data of JENDL-5 underestimate the independent fission mass yields around the peak of light fragments.
Since JENDL-5 did not take into account the effects of TKE and PFN and these correlations in its evaluated fission yield data~\cite{Tsubakihara2021}, this may trigger the deviations of independent fission mass yields around the peak of light fragments.
Figure~\ref{fig:rud54} shows the calculated IFY together with experimental data~\cite{1990ADNDT..45..239R}.
Since the number of free parameters increases, we can observe some improvements comparing with Fig.~\ref{fig:ind5}, for example, in Rb, Ag and In isotopes.
The present result of GP+GLS is comparable to the evaluated data of JENDL-5; however, we obtained some improvements for Ga, Rb, and In isotopes.
We also noticed underestimations of GP+GLS for Ba isotopes; however the cumulative yields of those nuclei are not deficient from the experimental data. 
For example, the present results of $^{144}$Ba and $^{145}$Ba are $4.20\%$ and $1.38\%$, respectively, while the experimental data are $4.40\pm0.66\%$ and $1.90\pm0.32\%$~\cite{Tipnis1998}, respectively.
\par
\begin{figure}[hbtp]
\centering
\includegraphics[width=0.98\linewidth, bb=0 0 620 480]{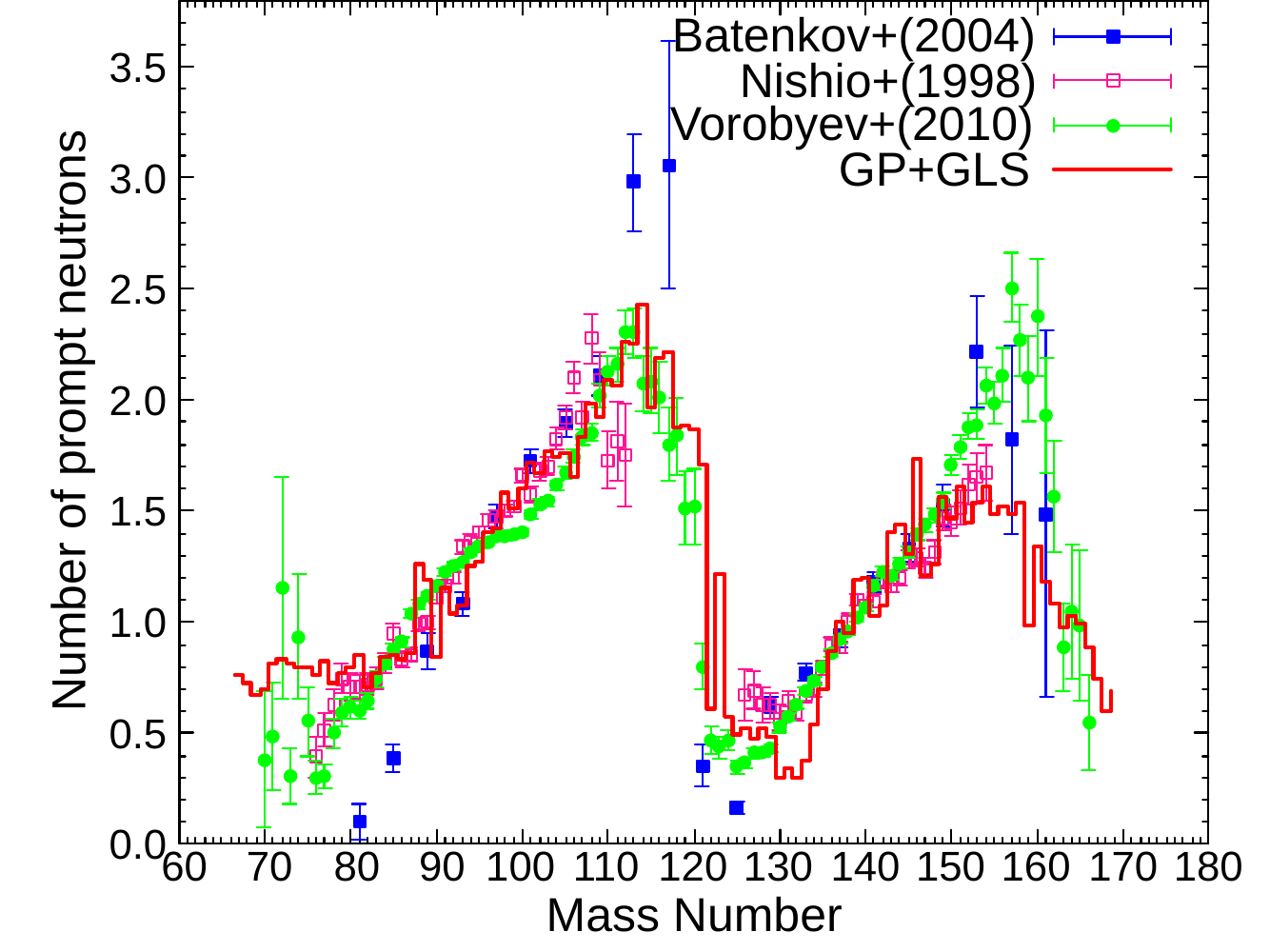}
\caption{Number of prompt neutron multiplicity as a function of fission fragment mass in case of full parameter set for $\delta Z_{p}$ and $\delta \sigma_{p}$.}
\label{fig:pnapga54}
\end{figure}

Figure~\ref{fig:pnapga54} is the result of the number of prompt neutron multiplicity as a function of fragment mass.
The saw-tooth structure observed in experimental data is nicely reproduced as we already observed in Fig.~\ref{fig:pnapga5}. 
The total number of prompt fission neutrons was $\nu_{p}=2.378$ for the case of 5 parameters for $Z_{p}(A)$ and $\sigma_{p}(A)$, while in this case  we obtain $\nu_{p}=2.407$ which is in a good agreement with experimental data and the JENDL-5 evaluation ($\nu_{p}=2.413$).
\par
\begin{figure}[hbtp]
\centering
\includegraphics[width=0.98\linewidth, bb=0 0 350 470]{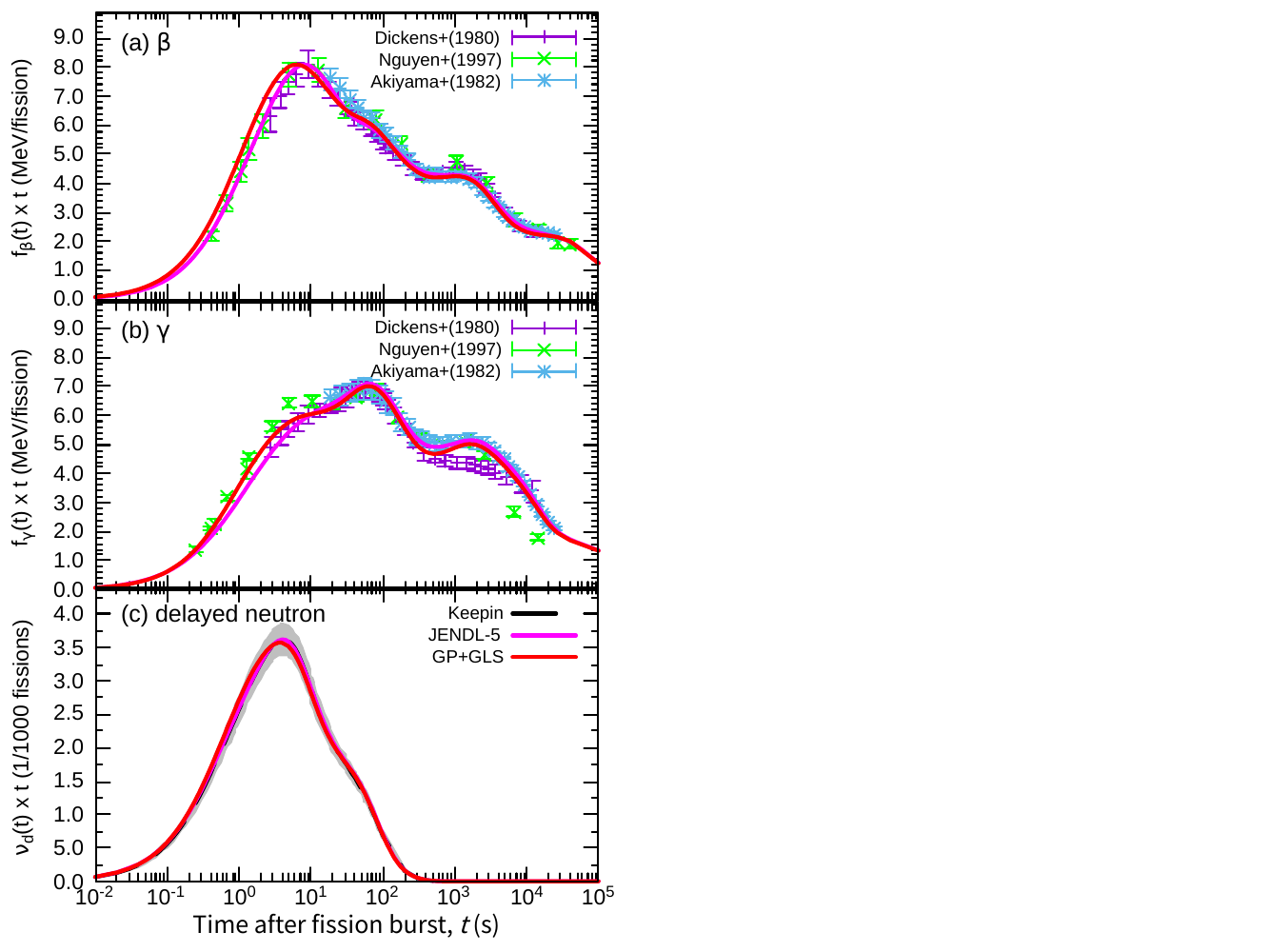}
\caption{Same as Fig.~\ref{fig:oyak5}, but for full parameter set for $\delta Z_{p}(A)$ and $\delta \sigma_{p}(A)$.}
\label{fig:oyak54}
\end{figure}

Figure~\ref{fig:oyak54} shows the decay heats of $\gamma$- and $\beta$-rays and delayed neutron yields.
The result of $\beta$-ray decay heats are slightly larger than the experimental data of Nguyen et al.~\cite{Nguyen1997} at $0.4\le t \le 5.0$ s, while it is close to JENDL-5 and nicely reproduces the experimental data after $t=5.0$ s.
Here, $t$ represents time after fission burst shown as $x$-axis in Fig.~\ref{fig:oyak54}.
The result of $\gamma$-ray decay heat is closer than JENDL-5 to the experimental data of Nguyen et al.~\cite{Nguyen1997} at $0.25 \le t \le 2.00$ s.
Since we have only 1 data set at the early stage ($t \le 2$ s) after fission burst, we are not still sure which data of the present work or JENDL-5 are more reliable both for $\beta$- and $\gamma$-decay heats, and experimental researches are thus required in this respect. 
Above $t=2$ s, both GP+GLS and JENDL-5 show a good agreement with experimental data of Akiyama et al.~\cite{Akiyama}.
Note that the data of Akiyama et al.~\cite{Akiyama} are those for fast neutron fission; however, we expect the result does not differ so much from thermal-neutron induced fission.
Next, we discuss the result of delayed neutron yields shown in Fig.~\ref{fig:oyak54}(c).
Comparing the result for the $5$ parameters for $\delta Z_{p}(A)$ and $\delta \sigma_{p}(A)$ in Fig.~\ref{fig:oyak5}, the calculated delayed neutron yields are drastically improved and comparable to the evaluated data of JENDL-5.
So, the parameter search with enough number of free $\delta Z_{p}(A)$ and $\delta \sigma_{p}(A)$ is important.
The number of total delayed neutron yields is about $\nu_{d}=0.01595$ for GP+GLS, which reasonably reproduces the experimental data of Keepin et al.~\cite{Keepin1957} ($\nu_{d}=0.0158\pm0.0005$).
\section{Summary}
\label{sect:summary}

We calculated fission fragment yields of the thermal-neutron induced fission of $^{236}$U and relevant fission observables with a newly developed CCONE system.
To determine parameters involved in this system, a combination of a Gaussian
process and a least square fitting was introduced.
At first, we demonstrated the performance of parameter search method of GLS and GP+GLS.
It was found that the GP+GLS method successfully found the most likely parameter set and was suitable for the present purpose as like involving a lot of parameters exceeding $100$.
We have shown that fission observables such as prompt fission neutrons, decay heats, and delayed-neutrons were reproduced in a comparable level to JENDL-5.
The total number of prompt neutrons and delayed neutrons were also reasonably reproduced. 
This fact indicates that the present approach has a potential to apply a practical evaluation of fission fragment yields.
\par
We still have a few questions arising from this work.
We could reproduce the overall structure of experimental data with only $5$ parameters for $\delta Z_{p}(A)$ and $\delta \sigma_{p}(A)$ except delayed neutrons.
This means that we may have a possibility to reproduce experimental data with less number of parameters than $54$ for them.
If it is true, an evaluation of fission fragment yields with this framework becomes more effective because we can reduce the computational time.
In addition, we observed that a staggering structure in $\delta Z_{p}(A)$ and $\delta \sigma_{p}(A)$ in Fig.~\ref{fig:Zp54}, which was not found in the Wahl's evaluation.
It will be interesting to confirm this structure within a microscopic framework.
\par
Our future perspective is to apply this framework to study different fission systems.
Now we calculate thermal-neutron induced fission of $^{233}$U and $^{239}$Pu.
We also plan to open the present result in the ENDF-6 format or other modern format to online so that anyone can access.
In this case, covariance data will be also provided as done in JENDL-5.

\section*{Acknowledgements}
The authors thank Dr. S.~Okumura at IAEA and Dr. T.~Kawano at LANL for useful comments. 
They also thank Research Coordinated Program.
FM acknowledges a special support from Dr. S. Yoshida at Utsunomiya university.
This work is supported by JSPS KAKENHI Grant Number 21H01856 and MEXT Innovative Nuclear Research and Development Program ”Fission product yields predicted by machine learning technique at unmeasured energies and its influence on reactor physics assessment” 
entrusted to the Tokyo Institute of Technology.
\bibliography{fpyccone}
\bibliographystyle{tfnlm}

\end{document}